%% file: main.tex
\title{Literature Survey}
\documentclass[journal]{IEEEtran}

\usepackage[space]{cite}
\usepackage{graphicx} 
\usepackage{epstopdf}
\usepackage{amsmath}
\usepackage{verbatim}
\usepackage{amsfonts}
\usepackage{amssymb}
\usepackage{floatrow}
\usepackage[tableposition=top]{caption}
\usepackage{caption}
\usepackage{siunitx}
\usepackage{adjustbox}
\usepackage{lscape}
\usepackage{rotating}
\usepackage{subfig}

\usepackage{tikz}

\renewcommand\IEEEkeywordsname{Keywords}
\newcommand\copyrighttext{%
  \footnotesize This work has been submitted to the IEEE for possible publication. Copyright may be transferred without notice, after which this version may no longer be accessible}
\newcommand\copyrightnotice{%
\begin{tikzpicture}[remember picture,overlay]
\node[anchor=south,yshift=10pt] at (current page.south) {\fbox{\parbox{\dimexpr\textwidth-\fboxsep-\fboxrule\relax}{\copyrighttext}}};
\end{tikzpicture}%
}

\begin{document}
\title{A Review of Computer Vision Methods in \\ Network Security}

\author{
	\IEEEauthorblockN{
		Jiawei Zhao,\IEEEauthorrefmark{1}
		Rahat Masood,\IEEEauthorrefmark{1}
		Suranga Seneviratne\IEEEauthorrefmark{1}
	}\\
	\IEEEauthorblockA{
		\IEEEauthorrefmark{1} The University of Sydney, Australia \\ Email: \{firstname.lastname\}@sydney.edu.au \\
	}
}

\maketitle
\copyrightnotice

\input{Abstract}
\input{Section_01}

\input{Section_02}

\input{Section_03}

\input{Section_04}
\input{Section_05}
\input{Section_06}
\input{Section_07}
\vspace{-2mm}
\input{Section_08}
\vspace{-2mm}
\bibliographystyle{IEEEtran}
\bibliography{main}

\vspace{-8mm}

\begin{IEEEbiography}[{\includegraphics[width=1in,height=1.25in,clip,keepaspectratio]{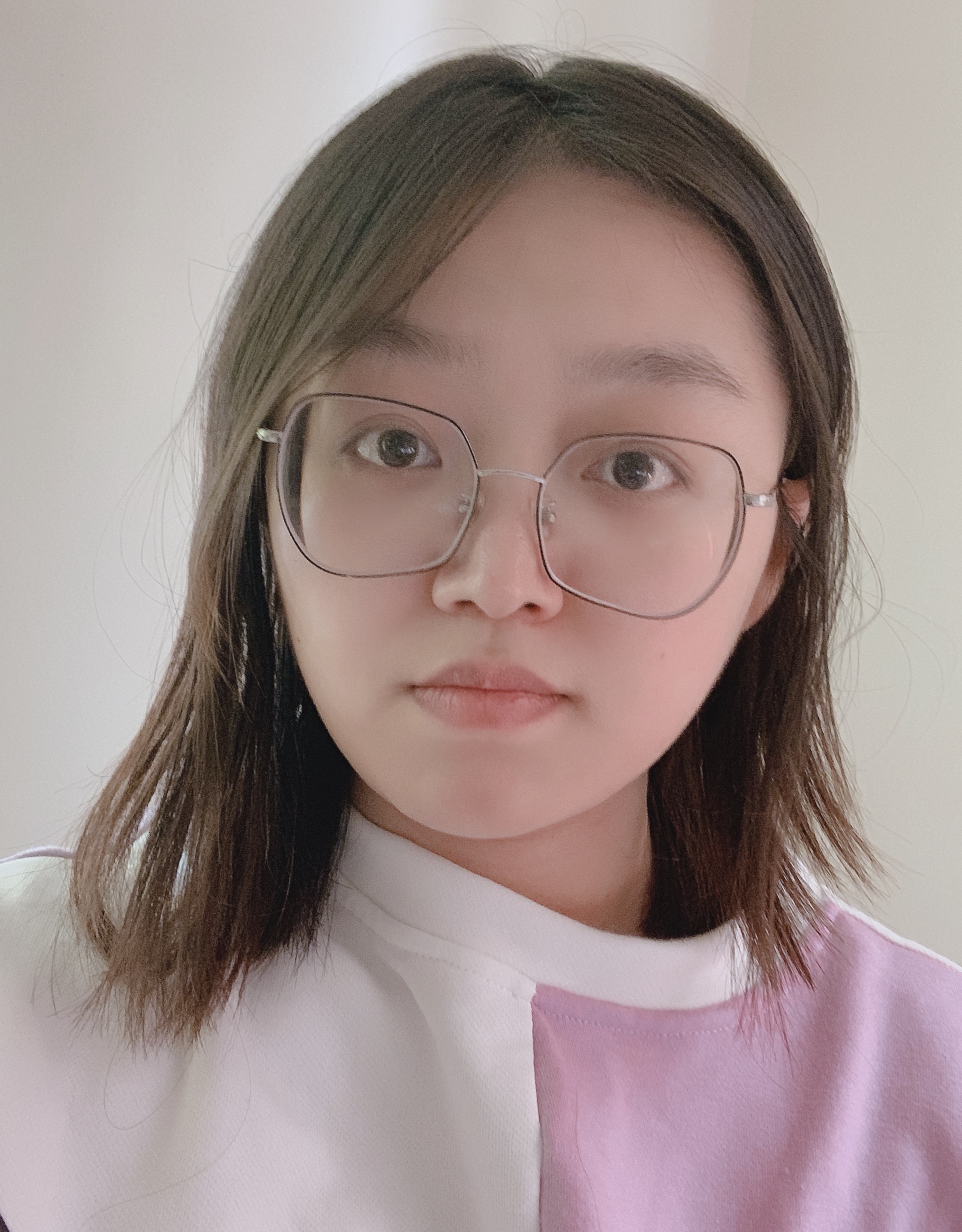}}]{Jiawei Zhao}

Jiawei Zhao received her Bachelor’s Degree in Electrical Engineering (First Class Hons.) jointly from The University of Sydney, Australia, and The Harbin Institute of Technology, China, in 2017. She is currently pursuing a PhD degree at The School of Computer Science in the University of Sydney, Australia. Her research interests include cybersecurity, deep learning, fraud detection, and security of machine learning.  \vspace{-15mm}


\end{IEEEbiography}

\begin{IEEEbiography}[{\includegraphics[width=1in,height=1.25in,clip,keepaspectratio]{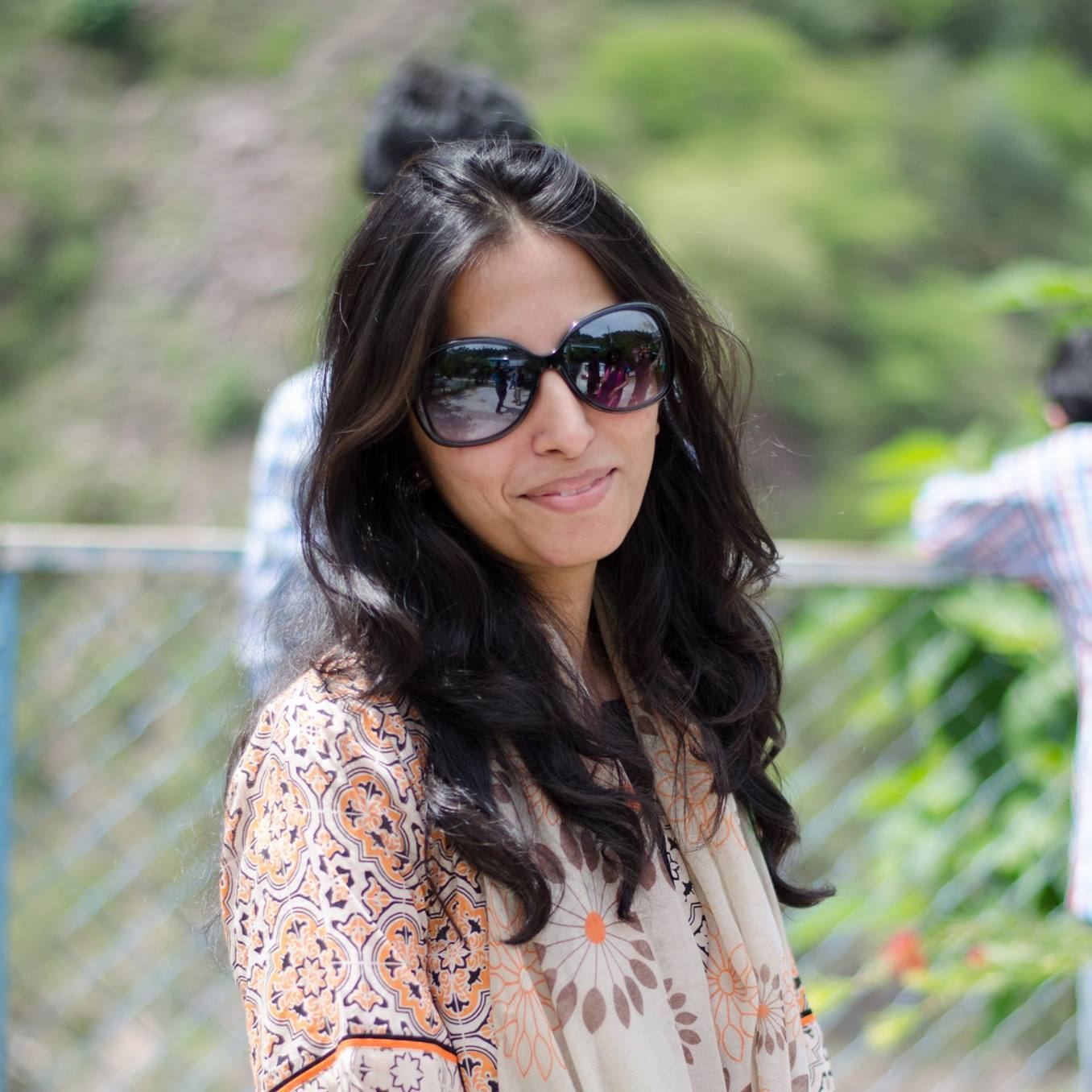}}]{Rahat Masood}

Rahat Masood is currently a research fellow at The University of Sydney. She received her PhD in 2019 from the University of New South Wales, Australia, in collaboration with information security and privacy group at Data61-CSIRO. Her research area focuses on security and privacy of mobile and web platforms, authentication and authorization, biometric security, and critical infrastructure protection. She was also a visiting scholar at the Sandia National Laboratories (SNL), New Mexico, and Cyber Security Policy and Research Institute (CSPRI) at The George Washington University. \vspace{-15mm} 

\end{IEEEbiography}

\begin{IEEEbiography}[{\includegraphics[width=1in,height=1.25in,clip,keepaspectratio]{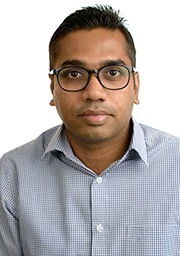}}]{Suranga Seneviratne}

Suranga Seneviratne is a Lecturer in Security at the School of Computer Science, The University of Sydney. He received his PhD from University of New South Wales, Australia in 2015. His current research interests include privacy and security in mobile systems, AI applications in security, and behaviour biometrics. Before moving into research, he worked nearly six years in the telecommunications industry in core network planning and operations. He received his bachelor degree from University of Moratuwa, Sri Lanka in 2005.

\end{IEEEbiography}

\vfill

\end{document}

%% file: Abstract.tex
\begin{abstract}
\label{abstract}
Network security has become an area of significant importance more than ever as highlighted by the eye-opening numbers of data breaches, attacks on critical infrastructure, and malware/ransomware/cryptojacker attacks that are reported almost every day. Increasingly, we are relying on networked infrastructure and with the advent of IoT, billions of devices will be connected to the internet, providing attackers with more opportunities to exploit. Traditional machine learning methods have been frequently used in the context of network security. However, such methods are more based on statistical features extracted from sources such as binaries, emails, and packet flows.  

On the other hand, recent years witnessed a phenomenal growth in computer vision mainly driven by the advances in the area of convolutional neural networks. At a glance, it is not trivial to see how computer vision methods are related to network security. Nonetheless, there is a significant amount of work that highlighted how methods from computer vision can be applied in network security for detecting attacks or building security solutions. In this paper, we provide a comprehensive survey of such work under three topics; \emph{i) phishing attempt detection, ii) malware detection, and iii) traffic anomaly detection}. Next, we review a set of such commercial products for which public information is available and explore how computer vision methods are effectively used in those products. Finally, we discuss existing research gaps and future research directions, especially focusing on how network security research community and the industry can leverage the exponential growth of computer vision methods to build much secure networked systems.

 \end{abstract}


%% file: Section_01.tex
\section{Introduction}\label{sec:introduction} 

We have been witnessing increasing trends in cyber attacks of various kinds. In 2018, there was a 133\% increase in the number of breached records compared to the previous year, and it was equivalent to a rate of 291 record breaches per second~\cite{BreachStats1}. It is estimated that in recent years, approximately 50\% of emails we receive are spam, and an average user receives 16 malicious spam emails per month~\cite{PhishStats}. Similar trends can be found for other types of attacks such as Distributed Denial of Service attacks (DDoS), ransomware, cryptojackers, and other malware. As such, network security has become an important aspect not only for large scale enterprises, but also for small and medium-scale institutions as well as for individuals. Security threats to networks are further expected to be increased with the deployments of IoT, where billions of devices are connecting to the internet and providing more opportunities for attackers. Thus, network security has become highly important more than ever it has been. For example, recent reports from Gartner, Inc.~\cite{gartner2019} forecast \$124 billion global spending in security, which is an 8.7\% increase compared to 2018. Regulations such as General Data Protection Regulation (GDPR)~\cite{gdpr2019} and security breach notification laws~\cite{BreachLaw1} are likely to further increase such spending over the coming years.

Recently, as a result of breakthroughs in machine learning, computer vision methods are evolving rapidly, in some cases, even surpassing the human-level performance. Such advancements are mostly becoming possible due to the wide availability of large volumes of labelled data and the processing advances in GPUs.  Traditionally, though machine learning methods are significantly used in network security and remains well-surveyed~\cite{buczak2015survey,gardiner2016security,xin2018machine,sommer2010outside}, use of computer vision methods for network security doesn't appear as a direct match apart from the obvious cases in authentication such as face, fingerprint, and iris recognition~\cite{zhao2003face,abate20072d,daugman2009iris,daugman2007new}. Here, we define computer vision methods in related to network security as methods using image representations of network data such as image feature representations (e.g. key point - descriptor methods such as SIFT~\cite{lowe2004distinctive} and SURF~\cite{bay2006surf}), various image hashing methods (e.g. perceptual hashing), and/or use of image-based machine learning methods such as Convolutional Neural Networks (CNNs) on data related to network security.

The trends to utilise advances in computer vision methods for network security is not solely an academic exercise. Recently the information security industry has also shown a keen interest in this area. For example, the latest RSA conference, which is a gathering of security academics and industry professionals, hosted RSAC Early Stage Expo that invited the industry to showcase their innovative projects, aiming to provide higher level of security for consumers and organizations worldwide~\cite{RSA}. Among the 50 identified promising start-ups, two starts-ups \textit{Pixm} and \textit{INKY}, were using state-of-the-art computer vision methods to detect phishing attacks in email inboxes and other channels such as Facebook, LinkedIn, and Instant messages~\cite{INKY, Pixm}. \textit{INKY} uses computer vision to block brand forgery emails and spear phishing attempts by spotting imposters by pixels. We provide more details on such products in Section~\ref{industry}. We notice that the industry focus to use computer vision techniques for improving security is quite a recent paradigm since most of the companies start introducing such tools from 2017~\cite{ind_foc1, ind_foc2}. 

In this survey, we explore the research carried out in the intersection between computer vision and network security. This type of a survey is essential as the field of computer vision is likely to continue advancing rapidly, and there is a unique, yet non-trivial opportunity to leverage such method to build security solutions for networked systems. Our survey is structured as follows. In Section~\ref{sec:taxanomy}, we explain the broader network security topics we have chosen and also describe the overall taxonomy of work under the selected topics. In Section~\ref{sec:PWA}, we survey the research carried out in the domain of \emph{phishing detection} using image matching methods. Section~\ref{sec:MDA} presents work where computer vision techniques are used in \emph{malware detection}, while Section~\ref{sec:ADA} surveys the work in \emph{traffic anomaly detection}. In Section~\ref{industry}, we survey some of the commercial products in this domain for which the public information is available. Section~\ref{sec:discussion} discusses the existing gaps in research, challenges to overcome, and potential future research directions and Section~\ref{sec:conclusion} concludes the paper.

%% file: Section_02.tex
\section{Computer Vision Methods in Network Security}
\label{sec:taxanomy}




As mentioned in the introduction, the applicability of computer vision methods in network security is not straight forward. We started our survey by referring to the network security threats taxonomy compiled by the  European Union Agency for Network and Information Security (ENISA)~\cite{enisa_1}. The report lists 15 types of threats in the likes of malware, web-based attacks, web application attacks, phishing, spam, denial-of-service, and botnets as the most prominent threats in recent times. Next, for each topic we searched academic publication repositories (e.g. Google Scholar, ACM Digital Library, and IEEE Explore) to check whether there are work that applied computer visions methods to detect such threats. During this process, we found several topics for which a body of work existed based on computer vision methods. Those topics included \emph{phishing detection, malware detection, traffic anomaly detection (targeting threats such as DDoS and BotNets), authentication solutions (especially biometric-based ones such as the face, iris, fingerprint recognition), and steganography}. Next, we short-listed network security topics based on their maturity in the field of computer vision as well as how important are they in the current network security threat landscape.

 
 Though authentication solutions are an essential element in network security, we noticed that authentication schemes based on face, iris, or fingerprints, have been significantly investigated and surveyed from the perspective of computer vision. Also, such schemes usually are straightforward applications of computer vision methods. For instance, Jafri et al.~\cite{jafri2009survey}, Abate et al.~\cite{abate20072d} and  Zhao et al.~\cite{zhao2003face} extensively surveyed face recognition techniques that utilise various computer vision methods. Similarly, other image-based biometric modalities such as iris recognition and fingerprint recognition remain extensively surveyed~\cite{daugman2009iris,daugman2007new}. Thus, we exclude such topics from our survey. We also investigated steganography and found that it is an interesting application of computer vision methods in network security. There are  multiple recent work in steganography that leverage deep generative models~\cite{hayes2017generating, volkhonskiy2017steganographic}. However, steganography is also a well-surveyed topic~\cite{ li2011survey, cheddad2010digital}.
 
Thus, after this analysis,  we selected \textit{Phishing Detection, Malware Detection,} and \textit{Traffic Anomaly Detection}, as topics of this survey. As we discussed later, phishing and malware are by far the two leading root causes behind many data breaches and have caused damages in the scale of billions of dollars. Similarly, the third topic, \textit{traffic anomaly detection} covers topics in the likes of DDoS detection and BotNet detection, which are also becoming a significant threats to internet services. We next define and describe each of these topics with emphasis on the importance of coming up with solutions to detect such threats.

\subsection{Phishing Detection}

Phishing is a social engineering attack that manipulates human trust to obtain confidential information about users such as passwords, credit card numbers, usernames, and personal data. The attack is often performed through instant messages or emails where the users are prompted with web pages that appear to be legitimate. The nature of the content of the message or the email will entice the users to enter their user names and passwords, bank account details, or other important personal information. This enables the possibility of a second, more devastating attack using the harvested information.

Phishing attacks are an ongoing problem in the internet. Attackers are using many sophisticated methods such that even the most tech-savvy users struggle to distinguish phishing attempts from legitimate web pages.
According to Phishlabs report, phishing attacks grew 40.9\% in 2018, while targeting 83.9\% of industries that offers financial, email, cloud, payment, and SaaS services~\cite{phislabs}. Overall, 98.9\% of these attacks target corporate users for credential theft or email scams.

Many high profile network breaches such as iCloud celebrity pictures, Sony hack, Bangladesh Bank heist, and eBay data breach have been traced back to attackers infiltrating the network using a well researched and crafted phishing attacks misguiding employees to enter their credentials. Overall, 90\% of the data breaches occurred globally in 2017 were results of phishing attacks. Recent reports also indicated that 75\% of the organizations faced at least one phishing attack in 2017, and the average click rate of the targets is an eye-opening 9\%. According to FBI's Internet Crime Complaint Center, Global Business Email Compromise (BEC) losses exceeded 12 billion US dollars in 2018. In 2017, these scams cost organizations almost 700 million dollars~\cite{fbi}.

It is essential to detect phishing attempts before users click a malicious URL to minimise damages. In this regard, there are two automated phishing attack detection methods: i) blacklist/whitelist-based methods and ii) text similarity-based methods~\cite{khonji2013phishing, 017phishing}. Blacklist-based methods attempt to keep a list of domain names or exact URLs to known phishing websites and alert users if they are trying to visit those links. However, phishing websites are highly dynamic, and the average lifetime of a phishing web page is only a few hours~\cite{zdnet}. Therefore in many cases, zero-hour phishing attacks usually easily bypass blacklist-based methods. Whitelist-based allows users to browse only a set of web pages that are deemed safe, which is not practical in many situations. Text similarity-based methods delve into semantics of the textual content, be it an email or a web page and try to decide whether it is a phishing attempt or not. Nonetheless, this method is likely to fail in future with the increased use of code obfuscation techniques. As the fundamental characteristic of a phishing page is it's visual similarity to a legitimate target page turning into computer vision method appears as a natural alternative to consider. In Section~\ref{sec:PWA}, we investigate such solutions that have been proposed for detecting phishing attempts.

\subsection{Malware Detection} 

Malware is a collective term used for various malicious software variants such as viruses, trojans, ransomware, and spyware. Malware typically consists of a code developed by an attacker to cause damages to data and systems or to gain unauthorised access to a network. Different types of malware follow different propagation methods to increase the reach. For example, some malware are delivered in the form of a link over email and require the users to click on the link while some malware spread through a malicious file download into a system and are self-triggered. 





According to a recent report~\cite{AV-Test}, from 2015 to March 2019, 413.73 million new malware programs have been reported, out of which 137.47 million have been reported in 2018 alone. The report also states that on average 279,545 new malicious programs are discovered every day. Malware can cause varying levels of damages from a single computer to national-scale infrastructure. For example, in 2017, Petya ransomware outbreak which started in Ukraine and subsequently spread across Western Europe, North America, and Australia caused losses over \$10 billion~\cite{ petya2017}.

Traditional malware analysis is usually based on either static code analysis (signature-based) or dynamic code analysis (behaviour-based)~\cite{gandotra2014malware, ye2017survey}. The signature derived from static analysis, is a short sequence of bytes unique to each known malware, which allows identifying newly encountered malicious files. This type of a detection method is less responsive to new malware and fails when codes are obfuscated. On the other hand, methods based on dynamic analysis (behaviour-based), can detect malware based on its run-time activities. These types of detection methods, however, give high false positive rates where benign programs are falsely classified as malicious programs. Additionally, dynamic code methods are computationally expensive and require a lot of resources for proper execution. To alleviate these problems, recently computer vision approaches have been proposed in the context of malware detection. For example, computer vision methods have been shown to be successful in analysing codes as images when they are obfuscated. Also, computer vision speedups detection process by providing a summarised picture of possible attacks through visual representation such as Treemaps and Thread graphs. In Section \ref{sec:MDA}, we survey such research work.




\subsection{Traffic Anomaly Detection}

A network traffic anomaly refers to a deviation from the usual traffic profile in a network. Unlike our first two topics, a traffic anomaly does not directly identify an attack. Rather it indicates that something unusual is happening in the network, which can be caused by a range of attacks such as DDoS, botnet activity, port scanning, or even because of some malware-infected nodes in the network. For instance, in September 2016, Mirai malware launched a DDoS attack on the popular security news web site/blog; \emph{Krebs on Security}, by generating more than 620$\;$ Gbps of network traffic~\cite{mirai}. The malware infected over 600,000 IoT devices, turning them into an army of remotely controlled bots. The reason for the use of a large number of IoT devices is to bypass anomaly detection systems which monitor the IP address of incoming requests and filters them if there are too many requests form a limited set of IP addresses. According to recent Kaspersky reports, 50\% of all DDoS attack led to a severe disruption of services~\cite{ano_tr} and there were cases where the attack continued for hundreds of hours.





\begin{figure*}[!t]
\centering
\label{fig:phished_web}
      \includegraphics[width=0.9\textwidth, keepaspectratio]{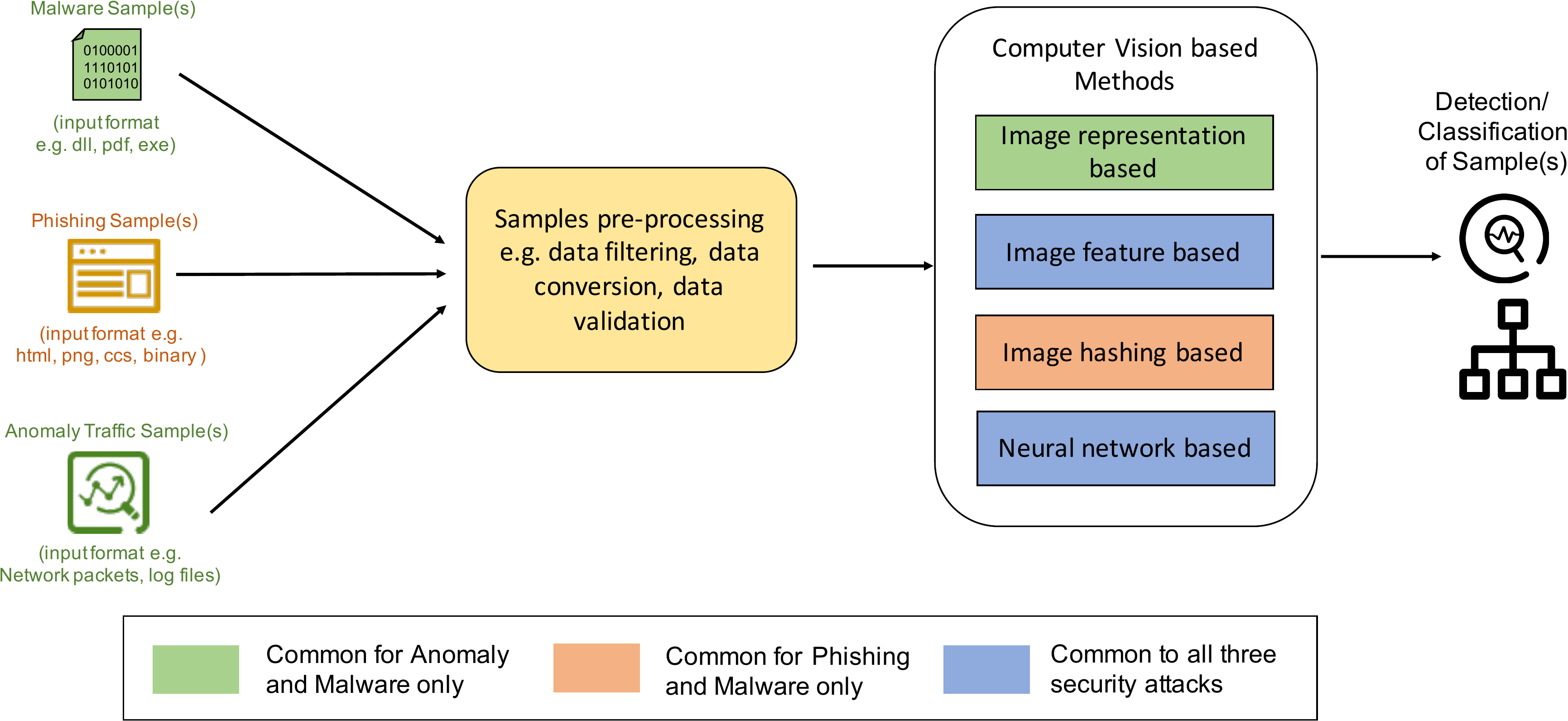}
\caption{Taxonomy of Computer Vision Methods in Network Security }
\label{fig:tax_overview}
\end{figure*}

Traditional network traffic anomaly detection approaches rely on underlying fitted models, which are labelled based on normal traffic behaviour. These approaches detect and characterise unusual patterns in the traffic based on deviations from the corresponding labelled data. Since these models are built on the knowledge of legitimate network traffic behaviour, the traditional detection approaches can identify previously unknown attacks. However, these conventional approaches suffer from high false positive rates. Also, if the set of rules (such as incoming IP addresses, port number, outgoing IP addresses, and protocols), defined within an underlying model are too complicated or too easy, then the accuracy of a detection system is affected. For example, an activity such as directory traversal on a targeted vulnerable server, which complies with a network protocol, easily goes unnoticed as it does not trigger any alert. Thus, efficiency of a detection system depends on how well rule set is implemented and tested on all protocols and scenarios. Another approach to detect traffic anomalies is based on computer vision methods, despite there are limited commonalities shared between network anomaly detection and computer vision tasks, such as image retrieval and object shape recognition. In Section \ref{sec:ADA}, we analysed a few traffic anomaly detection methods, that are based on computer vision.

\subsection{Structure of the Survey}

We structure our survey with separate sections to the three main topics that we survey. Across the three main topics we broadly categorise the computer vision methods that are being used in to four; \textit{i) image representation-based methods}, \textit{ii) image feature-based methods}, \textit{iii) image hashing-based methods}, \textit{iv) neural network-based methods} as described below and schematically summarised in Figure~\ref{fig:tax_overview}. However, we also highlight that we did not find work in all four sub-categories in all three main topics we survey.

\begin{itemize}

    \item \textbf{Image representation-based methods} have been applied in malware detection and traffic anomaly detection.  In these methods, a binary executable, log files, or a data file related to a malware or network traffic is first converted into a visual image and then visual fingerprints are generated based on malware behavioural activities such as API call sequences, I/O request packets (IRPs), resource consumption, network activities etc. These visual fingerprints are represented  through properties such as $x,y$ coordinates, RGB colours, and intensity level. Unlike the next described feature based methods that compare features, the image representation-based methods compare visual fingerprints to detect maliciousness. \\ \vspace{-1mm}
    
    
    \item \textbf{Image feature-based methods} generate features from image representations. For example, for phishing websites features are generated from the screenshots of the web pages or images/logos embedded to the web page. Example features include the traditional computer vision features such as  Scale Invariant Feature Transform (SIFT) features~\cite{lowe2004distinctive}, Speeded Up Robust Features (SURF)~\cite{bay2006surf}, and various image histogram features. Similarly, malware binaries or network traffic can be represented as images and features can be extracted. \\ \vspace{-1mm}
    
    \item \textbf{Image hashing-based methods} generate image representations that are efficient during retrieval and comparison tasks. For example, to identify visually similar web pages to a given page, a hash of the web page can be created. When a suspicious web page is found, the two hashes can be compared to see the new page is visually similar to the original page. Similarly, to improve the efficiency of malware detection approaches, local sensitive hashing (LSH) schemes have been introduced that checks the similarity between visual images of malware.   \\ \vspace{-1mm}
    
    \item \textbf{Neural network-based methods} accept some form of images as the input, yet do not require to generate features separately. For example, convolutional neural networks are representation learning method that do not require separate feature generation. The network itself learns what features are important to the task at hand. \\ \vspace{-1mm}
    
\end{itemize}

The rest of our survey is organised as follows. In Section~\ref{sec:PWA}, we provide detailed analysis on existing computer vision based phishing detection techniques, followed by malware detection and traffic anomaly detection in Section~\ref{sec:MDA} and~\ref{sec:ADA}, respectively. We then discuss industry tools or products in Section~\ref{industry}, that are utilising computer vision techniques for network security attack detection. In Section~\ref{sec:discussion}, we discuss open challenges in applying computer vision methods to network security, and also highlight possible future research directions. We conclude our survey in Section~\ref{sec:conclusion}.

%% file: Section_03.tex
\section{Visual Similarity based Phishing Detection}\label{sec:PWA} 

Many phishing attempts over various delivery forms such as emails, web, and mobile apps, try to misguide the target users by providing visually similar interfaces to the typical applications and services they use. As such, it makes sense to use computer vision methods in building phishing attempts detection systems. We found a significant body of research fitting into under three sub-topics we discussed in Section~\ref{sec:taxanomy}; {\bf \emph{i) image feature-based methods}}, {\bf \emph{ii) image hashing-based approaches}}, and {\bf \emph{iii) neural network-based approaches}}. All the work we surveyed applied some form of image transformation, and as a result, we do not report any work under the sub-topic of \emph{image representation-based methods}. \\ \vspace{-3mm}

\subsection{Image feature-based methods}
\label{SubSec:ImageFeature}

As mentioned before, image feature-based methods first identify an image representation of the possible phishing attempt. For example, the image can be a screenshot of the phishing web page. Then, the image is converted into a feature vector through various methods. Afterwards, different techniques such as $k$ nearest neighbour search, are used to find whether the web page is visually similar to a known legitimate page. The overall pipeline can be evaluated using performance metrics such as accuracy, precision, and recall using a ground-truth established dataset. We show this process in Figure 2. 

\begin{figure*}[!h]
\centering
\includegraphics[width=0.9\textwidth, keepaspectratio]{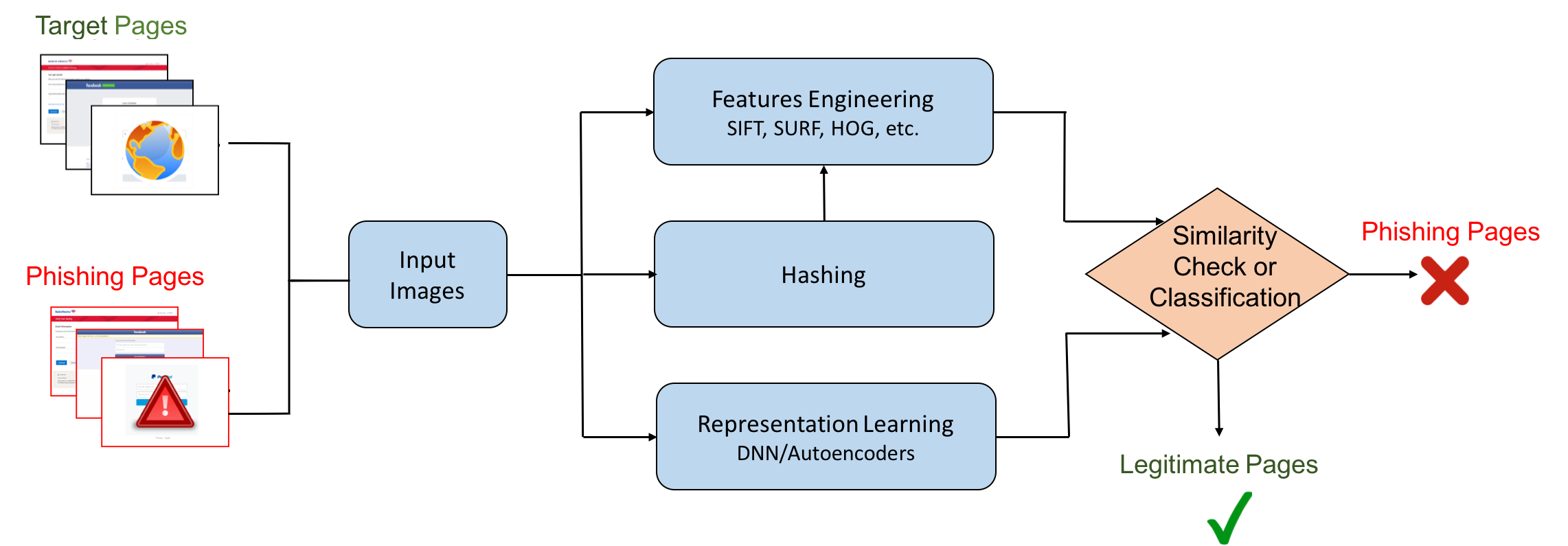}
\caption{Process of image feature-based methods}
\label{fig:phish_example}
\end{figure*}

As such the aspects to investigate under image feature-based methods for phishing detection are; {\bf \emph{input data (type of data, how authors collected the data, and how the authors established ground truth)}}, {\bf \emph{feature representation}}, {\bf \emph{similarity calculation}}, and {\bf \emph{performance levels}}. We next present work in this category with particular focus on above aspects. \\

\noindent{\bfseries \emph {i) Web page screenshot \& image content matching-based approaches}} \\ \vspace{-2mm}

One of the early work in this area is the SiteWatcher proposed by Liu et al.~\cite{PW:Liu2006,wenyin2005phishing}. It is service designed to run on mail servers and flag possible phishing links in both incoming and outgoing mail messages. All mails are monitored to check whether they contain suspicious words associated with a set of protected and sensitive web pages (potential targets for phishing attempts) for a given organization. If an email contains such words, SiteWatcher will consider all the URLs in that mail as suspicious and send those URLS together with the potential target URL to the \emph{visual similarity assessment module} for further analysis.     

The \emph{visual similarity assessment module} first identifies a set of \emph{salient blocks} by parsing the DOM (Document Object Model) tree of the web page. Salient blocks for a web page are the clearly distinguishable blocks such as header, footer, sidebars, and any other blocks in the page body that have consistent content either visually or semantically, and considerably different from other blocks~\cite{liu2004user}. Then, it considers three types of similarity between the blocks; \emph{block-level similarity, layout similarity,} and \emph{style similarity}. A set of features defines each similarity type. For instance, for \emph{block similarity}, \emph{visual similarity assessment module} first decides whether the block contains text of images and based on that calculate features for the block.  Text block features include features such as background colour, foreground colour, and anchor colour (colour of the hyperlinks). Examples of image block features are dominant colour of the image, image display, and image display height. The similarity between the two blocks in the phishing web page and the target is calculated as the weighted sum of the individual feature similarities, and the weights are empirically decided. Two blocks are considered to match if their similarity value is higher than an empirically determined threshold. Similar approaches are followed for other similarity types with different features.

Authors evaluated the SiteWatcher using a ground truth dataset that contained $320$ commercial bank home-pages and eight true phishing pages targeting six of the banks' pages based on the reports published by Anti-Phishing Working Group (APWG).\footnote{https://apwg.org} Authors considered the six true pages as the query to search for the visually similar pages in the dataset of 320 pages. If the similarity of any of the three metrics (block, layout, and style) were higher than a threshold $t$, the retrieved web page is reported as a potential phishing web page. Results indicated that for $t$ values greater than $0.7$ almost perfect true positive rate and false positive rate can be achieved. For a baseline comparison, authors also evaluated the performance of a pure text features based method and showed that such an approach have a much higher false positive rate.

While the paper has shown some promising results, it has several limitations. First, it identifies the target based on the presence of keywords and only activate the visual similarity assessment module if there are matches with a known set of keywords. Such a method is not scalable and may also miss more sophisticated attempts based on obfuscated code~\cite{kirda2006protecting}. Similarly, salient block identification is also text based DOM tree analysis which may not produce the optimal results with the use of client-side scripts and code obfuscation. The use of computer vision methods is limited to using manually curated image, colour, and style based features. Finally, the evaluation is limited to a smaller dataset, and it is unclear whether the methodology will scale to the range of thousands and hundreds of thousands of web pages and still able to maintain the same levels of true positive and false positive rates. Nonetheless, SiteWatcher is an important work in this domain as it proposes a useful architecture that allows building more effective and efficient phishing detection systems using computer vision methods. For example, salient block identification can be done entirely using computer vision algorithms and now by leveraging the recent advances of CNNs, we are in a position to design better similarity comparison methods.

Medvet et al.~\cite{PW:Medvet2008} proposed a phishing detection technique that considers the text pieces (including text style), embedded images, and overall visual appearance of the page to determine the page similarity. The DOM tree was used to compare the textual features such as content, colour, size and position. 2D Haar Wavelet transformation and colour histogram are the techniques used to analyze the embedded images and the overall image of the web page. Then a single similarity score is calculated and compared with a predefined threshold value to decide whether a web page is a phishing page or not. The authors claim that this method has zero false positive rate.

 The evaluation process contains two parts which are signature extraction and signature comparison. The comparison of the two signatures between two web pages requires multiple operations, corresponding to each feature. Authors found that it takes about 11 seconds to compare negative pairs and around 4 seconds to compare positive pairs. Authors also proposed a few optimizations, that significantly reduce the computational cost and make the time of negative comparison reduce to a few milliseconds. This solution is effective to detect phishing web pages that are visually similar to the legitimate ones. Though this work is inspired by AntiPhish\cite{PW:Engin2005} and DOMAntiPhish\cite{PW:Angelo2007}, compared to AntiPhish and DOMAntiPhish, it improves in the detection of phishing pages containing many embedded images. The approach is also similar to Liu et al.\cite{PW:Liu2006}, however has a much higher feature space.





In \cite{PW:FU2006}, authors proposed a solution that uses Earth Mover's Distance (EMD) to compare the visual similarity between web pages. In this method, the web page screenshots were first converted to a lower resolution of $100 \times 100$ using Lanczos resampling~\cite{turkowski1990filters} since it can generate sharp images. The features were calculated based on the colours of each pixel where a feature is the colour and the centroid of its position distribution in the image in a reduced colour space of 4,096 colours. Frequency of the colour (i.e. the total number of pixels using that colour in the image) was used as the feature weights to create the full signature of the web page. Finally, authors decided whether a new page is a phishing web page or not based on whether the EMD between two web pages is less than an empirically decided threshold. Authors conducted multiple experiments to evaluate different strategies of choosing the threshold.

To evaluate the proposed solutions, authors established a web page image dataset that contain images of 10,272 legitimate pages and nine phishing pages, which authors collected based on real-world phishing attacks. At the best threshold value, authors were able to identify eight out of nine phishing web pages with only 73 false positives, corresponding to a \emph{true positive rate} of 88.89\% and a \emph{false positive rate} of 0.71\%. Also, authors compared their approach with a completely text based approach and the above discussed region based approach~\cite{PW:Liu2006,wenyin2005phishing} and showed that new method have lesser \emph{false positive rate}. Work of ~\cite{PW:Zhang2011} et al. also used a similar EMD based approach.

This type of work are an improvement compared to Liu et al.~\cite{PW:Liu2006,wenyin2005phishing} and has proposed better methods of capturing the visual similarity as indicated by the results. Nonetheless, going to a lower resolution might lose some information when it comes to larger datasets. Use of only colour features can also be a limitation, and a better approach may be to use both colour and region-based features. Another limitation is that the use of EMD is not justified with results. It is unclear whether a distance metric like $L_2$ distance which is having lesser computation requirements can provide the same level of performance. Finally, the dataset had only nine cases of phishing attempts. Evaluation in a much larger dataset is required to get a better understanding of the final solution.

Chen et al.~\cite{PW:Chen2009} also started with the screenshots of target and phishing pages, but proposed to use more advanced methods of image representations. Authors first converted the web page image to grey-scale by averaging the red, green, and blue values of each pixel in the image. Then Harris Laplacian corner detection~\cite{mikolajczyk2001indexing} was used to identify a set of key points in the image that are usually invariant to  modifications in the likes of shifting, lighting variation, and colour transformation. To describe each key point, authors use a lightweight version of Contrast Context Histogram (CCH)~\cite{huang2006contrast}, which captures the contrast distribution in the neighbourhood of a key point. To make CCH computation lightweight, author only considered the grey-scale image of the web page. Authors calculated the similarity between two pages using Euclidean distance-based key point matching. The proposed solution was evaluated in dataset comparable to previously discussed results and achieved similar performance levels. 

Corona et al.~\cite{PW:Corona2017} developed a system called DeltaPhish to detect phishing web pages in compromised websites, which is the first system utilizing websites vulnerability for phishing detection. The overall process of DeltaPhish involves browser automation, feature extraction, HTML-based classification, snapshot-based classification, and fusion classification.

In snapshot-based classification part, the feature representations used to classify snapshots are HOGs (Histogram of Oriented Gradients) and colour histograms. Compared to other feature representation like SIFT (Scale-Invariant Feature Transform), the features used in this paper can achieve better performance for image classification of high inter-class similarity. Authors also do image tiling to gain more spatial information of snapshots. Thus, the features are extracted from the whole snapshot picture, its quarters and sixteens. Based on the feature vectors, the visual similarity is evaluated for input image and homepage. Finally, authors leverage a linear SVM to do the classification. The true positive rate of only snapshot-based classifier is about 82.5\% at 1\% false positive rate. The processing time of snapshot-based classifier is more than 1.2 seconds averagely, which HOG feature extraction takes most of the time, and colour feature extraction only takes less than 3 ms averagely. The limitation of this method is that it depends on the assumption that within a website, the legitimate pages have a similar visual appearance with the homepage.

In a slightly different approach of using the web page screenshots, Lam et al.~\cite{PW:Lam2009} started by identifying non-overlapping regions called \emph{layout blocks}. This was done by first using Otsu's threshold methods to convert the screenshot to black and white and identifying the blobs that have the same colour. Then these blobs were used as a mask to identify regions in the original image. The similarity between two pages was calculated as the sum of the differences in width, height and location of each pair of blocks. Then authors used empirically established thresholds to decide what is the minimum similarity required to decide on a phishing web page. Compared to the EMD algorithm used in~\cite{PW:Liu2006}, this method is faster and can handle non-square image inputs.

Different from the previous papers, GoldPhish proposed by Dunlop et al.~\cite{PW:Dunlop2010}, used optical character recognition (OCR) to extract text from a screenshot of a web page and used that text to decide whether the page is a phishing page or not. Authors, first take a $1200 \times 400$ pixel image of the web page, covert it to black and white, and extract the text of the image using a commercial OCR software. As the image also contains the logos present in the web page, the OCR will extract any text available in logos, which authors claim as an advantage compared to other methods. The extracted text is entered into Google Search as the queries, and related websites were retrieved. The intuition here is that, as the text in logos, and the text in websites are indexed by Goolge, the search will return the corresponding legitimate site (if there is any) as one of its top results.  

Testing GoldPhish on 100 active phishing web pages identified from PhishTank and 100 legitimate web pages, showed good results with only two false positives and zero false negatives. Authors also compared their method with text-based methods and heuristics-based methods and showed that GoldPhish achieves better results.

While the results of this method are promising, it has a significant limitation. Use of OCR can be justified only for the websites that have a logo at the top of the web page, and the logo has text in it. For logos, that do not have embedded text, Goldphish might perform worse than logo detection-based approaches. Otherwise, Goldphish is expected to perform the same as a solution that does text matching between web pages without any image processing methods. Also, the comparison was made only with simple methods, and thus it is difficult to compare how GoldPhish will perform compared to other image-based approaches. Nonetheless, the idea of using the Google indexing system as we discussed earlier under~\cite{PW:Fatt2014} is innovative. Due to the limited lifetime of phishing web pages, they are unlikely to be indexed by Google. Thus, the text queries will get hits from the legitimate web page.

In a recent work, Tian et al.~\cite{PW:Tian2018NeedleIA} also highlighted the necessity of adding visual features to the phishing detection problem to overcome the content obfuscation methods used by attackers. Authors used the open-source OCR engine Tesseract\footnote{https://github.com/tesseract-ocr/} that used the detected words as features in data going into a machine learning classifier.

Hara et al.~\cite{PW:Hara2009} enhanced phishing detection based on image similarity by detecting phishing sites even without the original registered sites. The comparison of two sites in this method is not only between phishing sites and their victim sites, but also between phishing sites that spoof the same site. The reasons why this method does not need an initial database are: 1) many phishing sites mimicking a same site are visually similar to each other;  2) their system will discover sites on which there is a different image shown from the previous one and use these sites to detect new phishing sites.

By taking the screenshot of web page, Hara et al. utilised an image searching tool called ImgSeek~\footnote{http://www.imgseek.net/}, which can output images similar to the input image and their similarity score. Two hundred twenty-four original legitimate sites spoofed by 2262 phishing sites were extracted from 521 legitimate sites by their system. The detection rate is 82.6\% while the false positive rate is 18.0\%, which can be improved by adding whitelist. Even though this method has a high false positive rate, it can handle many kinds of phishing sites and combine with other methods. Notably, compared to Liu and Medved~\cite{PW:Medvet2008, PW:Liu2006}, this system does not rely on an initial set of legitimate pages. \\


\noindent{\bfseries \emph {ii) Logo detection-based approaches}} \\ \vspace{-2mm}

Wang et al.~\cite{PW:Wang2011} proposed a slightly different approach by using the logos embedded in web pages than using the entire screenshots. The method consists of three steps: first, authors determine if the web page contains the given logo. Second, they check if the brand owner has authorised host an IP address to use its logo. Finally, for sites containing unauthorised logos, if the user enters keyboard input on the page, the system will issue a warning.

For the first part, authors use a simple heuristic approach to identify possible logos on the web page under the assumption that logos are usually placed at the top of the web page. Starting from the top of the page, stripes of height $N=100$ are considered, and continuous blocks of single colour are removed as those parts are unlikely to contain logos. For the remaining blocks, SIFT representation~\cite{lowe2004distinctive}, which is another key point and descriptor-based representation similar to above mentioned CCH, is calculated and matched against a local copy of a logo image database. If there is a match, defined by having a Euclidean distance less than a threshold, the brand owner is contacted to check whether the use of logo in the current web page is allowed using two methods; a modified DNS query or certificate base approach where the original logo contains the brand owners public key.

To measure the performance of the solutions, authors first established a dataset of popular logos that covered 166 distinct brands (again taken from 176 distinct instances. The dataset was augmented with the colour inverted versions as it was known to a common technique used by the phishing attempts (i.e. 352 logos). The second dataset was the "brand dataset" that contained screenshots from 23 legitimate websites that are commonly targeted by phishing attacks. For each website, authors collected ten pages, and thus the total size of the \emph{brand dataset} is 230. The third dataset contained the 219 screenshots of known phishing sites taken from PhishTank database.\footnote{https://www.phishtank.com/} Authors were able to achieve a true positive rate of 90\% at 3\% false positive rate. 

While the use of SIFT might improve the image matching, this work has a significant limitation in detecting logos, which is done based on a straightforward heuristic algorithm. Such an approach is likely to be inefficient in cases where the web site banner does not have large areas of a single colour. In such a setting, the area the algorithm selects, may not necessarily be logos and might be quite large, resulting in a high latency when it comes to SIFT key point calculation. While it is understandable, that when the authors did the work, object detection methods were not evolved and now this work can be improved by leveraging the recently proposed deep learning-based logo detection methods~\cite{fehervari2019scalable,wang2018deep}. Another limitation is the latency. The solution is proposed as a safe browsing solution in contrast to the above discussed email solutions. When it comes to browsing, user experience is highly important, and users will be annoyed by slightest delays. Thus, the total latency (consisting of logo identification, SIFT key point and descriptor generation, and finally matching) in the range of tens of seconds is not acceptable. Nonetheless, such delays may suit well for an email security solution, where there is some lag between the mail is being received by the server and accessed by the user. Also, email server can control access to the email until the scanning is finished.





In \cite{PW:Zhou2014}, Zhou et al. proposed a method that combines logo detection and web page snapshot comparison. The \emph{logo detection and comparison} were based on SURF (Speeded Up Robust Features)~\cite{bay2006surf} descriptors which are faster to calculate than previously discussed SIFT key points and descriptors. For each suspected web page screenshot and each logo image, the authors identified two sets of interest points using the SURF detectors and then generated two sets of SURF descriptors. For each descriptor of the logo image, the descriptor that matches best (above a defined threshold) in the suspected web page is identified using Euclidean distance. Nonetheless, still, there are matches between the logo and irrelevant parts of the web page. To address this, authors proposed to use homography transformation~\cite{chum2005geometric} estimated using random sample consensus (RANSAC)~\cite{fischler1981random}. Homography transformation allows projecting the left top and right bottom coordinates of the logo image to the suspected web page image. By doing so, authors then make some decisions based on heuristics such as whether the two points are apart enough to be a logo or which part of the web page the identified logo is located. For \emph{snapshot comparison} authors used the same approach as Fu et al.~\cite{PW:FU2006} that was discussed earlier.

Authors collected three types of data to evaluate the performance of their system. The \emph{phishing web page} dataset contained 2,129 phishing web pages that were collected from PhishTank. As \emph{target web page dataset,} authors use the screenshots of the web pages of PayPal, PosteItaliane, eBay, iTunesConnect, and MyAppleID. Finally, as the \emph{irrelevant legitimate web page subset}, authors collected snapshots of 1,367 web pages by discovering URLs via Google search for 26 keyword searches. Overall, the authors were able to achieve ~91\% true positive rate and ~1.46\% false positive rate.

While on paper, it can be expected that combination of the ideas of logo comparison and snapshot comparison should improve the detection rates of phishing web pages, the results of this paper do not justify that there is extra benefit. Nonetheless, all the solutions are using different datasets, and it is difficult to establish which methods are superiors. This leads to the following discussions we make on the requirement of establishing benchmark datasets in this domain. Also, as mentioned above, we emphasise that significant improvements have happened in computer vision space in the last few years and those ideas can be borrowed to improve this type of work (e.g. logo detection and comparison).

PhishZoo~\cite{PW:Afroz2011} used logo matching in a much broader phishing detection systems that consider other similarities between web pages in the likes of URL patterns and HTML text. The image similarity aspect of the solution involved keeping a logo database and matching all the images content of suspected web pages against the stored logos using SIFT key points. Performance evaluation showed that using text features only resulted in very high accuracy (97.6\%) but also a very high false positive rate (18.7\%), which is not desirable. On the other hand, use of only image matching resulted in a moderate accuracy (82.7\%), but a very low false positive rate (2.5\%). Overall the combination of images and text resulted in an accuracy of 90.2\% and 0.5\%. 

This work provides an important justification on why image-based methods are required to be incorporated into usual text-based phishing detection systems. Nonetheless, authors highlighted some notable limitations they identified throughout their work. First, the SIFT image matching can identify only near similar matches. For instance, if the logo is rotated significantly, cropped, or combined with other elements, SIFT fails to give a high score. Second, the latency of the system, which was in the range of 7-17 seconds, is mainly contributed by the image matching phase, than the SIFT key point generation (i.e. based on the number of stored logos). As a result, more efficient signature matching methods need to be used in real-world deployments. 



In 2014, Fatt et al.~\cite{PW:Fatt2014} proposed a method based on website \emph{Favicons} (i.e website shortcut icon displayed on the address bar of a web browser) to do phishing detection by utilizing the Google Search-by-Image API. Then they used the results of the Google search engine to evaluate the authenticity of the website. Intuition of this approach is that, if the \emph{Favicons} is a copy of a popular website or a popular logo, Google search results will list the target website first as well as some texts describing the logo (e.g. \emph{Best guess for this image: icon paypal}). 
Then based on the solutions returned by the search query, the authors used four text-based heuristic rules to decide whether the website from where they got the \emph{Favicons} is a phishing site or not. Authors used 500 phishing pages (chosen from PhishTank) and 500 legitimate pages (chosen from Alexa top 500 global websites) to validate the performance of their approach. Authors reported a 97.2\% true positive rate and a 2.8\% false positive rate. 

In contrast to previous methods that keep their own images of the target pages, this method leverages the Google's page indexing system which is an advantage as it eliminates the need of maintaining a database of target web sites. As indicated by results, this method will provide very high accuracy for target pages that are indexed by Google and sufficiently widespread. Nonetheless, the major drawback is that it will work only if the \emph{Favicons} is available on the web page. As a result, this is a method more suitable to be one specific feature of much larger solution covering various other aspects.

\subsection{Image hashing-based approaches}

One common aspect of all the work discussed in above Section~\ref{SubSec:ImageFeature} is that once the feature representation of a suspected page is generated, a \emph{$k-NN$} ($k$ nearest neighbour) search is required to identify the target page from a set of potential pages stored in a database. This search can be time-consuming as highlighted by some work~\cite{PW:Afroz2011} and is proportional to the size of the target web page database. This is exacerbated by the higher complexities in calculating EMD and SIFT representation. Though this search process can be optimised by using more efficient data structures such as $k-d$ trees and \emph{vantage-point trees}~\cite{kumar2008good}, there is a fundamental requirement of making it more efficient. We next discuss a set of work, that addressed this problem by using \emph{image hashing methods}. 


In~\cite{PW:White2012}, the phishing websites were detected by analyzing their structural characteristics and page screenshots. The primary computer vision technique used by the authors here was to compute a hash of the web page screenshot and retrieve similar images based on the Hamming Distance. Authors experimented with three hashing methods; \emph{MD5}, \emph{SHA-512}, and \emph{pHash (Perceptual Hashing)}, in which latter is specifically designed for image matching while the former two are more suitable for string/text content matching. Authors first demonstrated the performance benefits of their method using a limited dataset of five phishing pages taken from PhishTank. As expected, \emph{pHash}, which is specifically designed for image matching performed better compared to \emph{MD-5} and \emph{SHA-512} that are more suitable for string or text content matching.

Then, the authors conducted a much large scale study by collecting URLs from Tweets and checking whether there are phishing ULRs or not using the proposed approach. Authors collected 1,829,531 URLs from Tweets. Authors found over 200 phishing pages trying to impersonate top phishing targets. The average processing time to decide on a page was less than 4 seconds, which was an improvement compared to previous work.

This work demonstrates the efficiency of hashing based image matching methods. Nonetheless, there are many other image hashing methods such as \emph{block hashing, wavelet hashing, } and \emph{difference hashing} and more thorough experimentation is required to identify which methods works well for website screenshot matching and provides a good trade-off between accuracy and computational efficiency.

Fotiou et al. designed a model to detect phishing websites based on the perceptual hash of the web pages' screenshots~\cite{PW:Fotiou2012}. The whole mechanism contains a web browser extension, a back-end application and a database. The web browser extension is responsible for capturing a screenshot of the currently loaded web page. The back-end application is used to extract and compute information of page screenshots, and the database is used to store all information signatures. In their implementation, authors used the open-source perceptual hash library  \emph{(pHash)}, and they evaluated information signatures by applying perceptual hash over a web page's screenshot bytes. And then authors used the hamming distance of the hash value to compare the similarity of the screenshots. Similarity threshold represents the maximum distance between two almost identical hash values of screenshots. If the numerical distance of the hash value of two screenshots is greater than the similarity threshold, it will be considered as a phishing attempt. 

The dataset contains 100 phishing pages captured from PhishTank database and 100 legitimate pages obtained from the list of  Google’s top 100 most visited sites in the United States. As a result, by choosing a good similarity threshold, the authors could achieve a recall of 81\% and false positive rate as low as 1\%. Even though the dataset is small, the result still shows a real-time performance. While this work is similar to~\cite{PW:White2012}, no comparison was provided.


\subsection{Neural networks-based approaches}
\label{sub:phish_neural}
One of the recent advances in computer vision is CNNs and other related deep learning methods such as auto-encoders that are able to eliminate the need of curating features manually; similar to what was done in above discussed work that were using SIFT or SURF. As such, we next investigate work that leverage the representation learning capabilities of deep learning methods to detect phishing web pages. As phishing detection work spans beyond the point where CNNs became mainstream (i.e. after the success of AlexNet~\cite{krizhevsky2012imagenet} in the ImageNet competition in 2012), we also found some work that used traditional neural networks, and we discuss that work here as well.

Borgolte et al.~\cite{PW:Borgolte2015} proposed MEERKAT to detect \emph{de-faced websites}, which is a different but potentially a related problem to phishing web page detection. In a \emph{de-faced} or a \emph{vandalised} website, an attacker replaces some content of the legitimate site, and as a result it's visual appearance is different from what it is supposed to be. First, authors extract $160 \times 160$ windows from web page screenshots for both de-faced and legitimate web pages. Then, authors build a deep neural network to learn advanced features from these windows automatically. The structure of this deep neural network is comprised of a stacked auto-encoder and a standard feed-forward neural network. The stacked auto-encoder is utilised to de-noise the images and learn high-level features from the images. The feed-forward neural network with dropouts is leveraged to do the final classification. 

Authors showed the feasibility of MEERKAT on a large website defacement dataset that included 10,053,772 damages observed between January 1998 and May 2014, and 2,554,905 legitimate websites. Overall, MEERKAT achieved a true positive rate between 97.422\% and 98.816\%, a false positive rate between 0.547\% and 1.528\%, and a Bayesian detection rate between 98.583\% and 99.845\%.

While the problem that this work addresses is different, the same approach can be used for phishing web page detection. The idea of learning latent representations from data itself, rather than manually identifying and calculating features will improve the efficiency of designing a phishing detection pipeline. The method is right now based on windows; therefore further work is required to understand what is the suitable setting for phishing web pages. For example, the model can take the whole screenshot as the input, only some key segments, or random patches and such decisions require further analysis.

Adebowale at el.~\cite{PW:Adebowale2018} proposed ANFIS (Adaptive Neuro-Fuzzy Inference System), a traditional neural network and fuzzy logic system based on calculated feature representations. Authors created 35 features from web pages that include 22 text-based features to represent the structure of the page, eight features to represent the frame-based properties, and five features to represent the image resources of the web page. However, the image features are not directly related to the visual similarity or the content of the images. For example, \emph{favicon} feature indicates whether the icon is loaded from a domain other than the domain shown in the address bar. As such, this work can't be considered as a working that truly uses computer vision methods for phishing detection.  

In 2019, Abdelnabi et al.~\cite{PW:abdelnabi2019} proposed a new visual similarity-based framework, called WhiteNet, utilising a triplet convolutional neural network for phishing website detection. Authors also released a new dataset called WhitePhish, which can be used to further benchmark phishing detection methods. The dataset contains phishing web pages, legitimate training web pages, and legitimate test web pages. The triplet learning process that was successfully first used in face recognition~\cite{DBLP:journals/corr/SchroffKP15} involves three inputs; an anchor image, a positive image which is a screenshot of the same website as the anchor and a negative image from the different website. The triplet loss that is being optimised during the training process attempts to make the embedding of samples in different classes further apart from the embedding of samples in the same class. Authors were able to achieve a recall of  95.81\% at a FP rate of 6.88\%. This work indeed is an example of how modern computer vision ideas can be utilised effectively to phishing detection.

\begin{table*}[t!]
\tiny
\caption{Summary of Visual Similarity Based Phishing Detection Approaches }
\label{Tab:PhishSummary}
\begin{tabular}{p{2.0cm}|p{2.4cm}| p{2.0cm}|p{1.3cm}|p{5.3cm}|p{2.7cm}}

 \hline
\textbf{Research Work}	& \textbf{Computer Vision \newline Methods} 	& \textbf{Input Image \newline Structure} & \textbf{Dataset Size (Phishing/ Legitimate)} & \textbf{Summary} & \textbf{Performance Metrics	}\\ \hline

\hline
\multicolumn{5}{c}{\textbf{Web Page Feature Representation Approaches}} \\ \hline
\hline
   
   Liu et al.~\cite{PW:Liu2006} & Visual Similarity Assessment based on Key Regions, Page Layouts, and Styles & N/A &8/320   & Compares the suspect page to the legitimate page by measuring block-level similarity, layout similarity, and overall style similarity. & Recall - 100\% \newline Precision - 91.2\% \\ \hline 
   
   Medvet et al.~\cite{PW:Medvet2008} & Visual Similarity based on 2D Haar Wavelet Transformation and Colour Histograms & N/A	& 41/161 & Extracts three visual page features which are texts, embedded images, and overall images of the from the snapshot of a web page and then computes the distance between the colour histograms, the positions and the 2D Haar wavelet transformations to gain a similarity score. & Recall - 92.6\% \newline False Negative Rate- 7.4\% \newline False Positive Rate - 0\% \\ \hline

    Lam et al.~\cite{PW:Lam2009} & Layout Block Matching  & N/A  & 6,750/ \newline  312	& Converts both the suspicious page and the legitimate page to black-and-white images, divides images into different layout blocks, and then performs layout matching between them to get their similarity score. & Accuracy - 99.6\% \newline False Positive Rate - 0.028\% \\ \hline
   
    Fu et al.~\cite{PW:FU2006} & EMD-based Visual Similarity calculated on Colour and Coordinate Features	& 100*100 & 9/10,272  & Converts the screenshot of web pages into low resolution images with normalised size, and then uses EMD to assess the visual similarity based on colour and coordinate features. & Recall - 88.8\% \newline Precision - 99.7\% \\ \hline 
   
   Zhang et al.~\cite{PW:Zhang2011}	& EMD-based Visual Similarity calculated on Colour and Coordinate Features	&100*100 &5,711/\newline 10,272 & Transforms web pages to images, generates image signatures and measures the distance of images by EMD, and finally compares the visual similarity with the threshold. & False Negative Rate - 0.88\%  \newline Classification ratio - 99.9\% \\\hline
   
   Chen et al.~\cite{PW:Chen2009} & Contrast Context Histogram (CCH)  & N/A &-/300 & Captures a webpage screenshot and extracts its keypoint feature, then uses CCH descriptors to perform image matching and finally classifies a page based on page similarity degree. & Accuracy - 95~98\%\newline False Positive Rate - 1\% \newline False Negative Rate - 1\% \\ \hline
    
    Corona et al.~\cite{PW:Corona2017} & Histogram of Oriented Gradients (HOGs) and Colour Histograms  & N/A & 1,012/\newline 4,499 & Captures a screenshot of a browser, then extracts visual features based on HOGs and colour histograms, and finally performed classification on the similarity between two histograms based on SVM. & Recall - 99\%, \newline False Positive Rate- 1\%\newline  False Negative Rate - 1\% \\ \hline
       
    Dunlop et al.~\cite{PW:Dunlop2010}	& OCR and Google Search API & 1,200*400 & 100/100 & Captures a page screenshot, and utilises OCR software to extract its text information, then searches the texts on Google to retrieve results. & Recall - 98\% \newline False Positive Rate - 0\% \\ \hline
    
    Tian et al.~\cite{PW:Tian2018NeedleIA}	& OCR and Image Hash & N/A &1,224/\newline  65,7663 	& Utilises OCR to extract texts from the screenshots of web pages as OCR features, and applies the Image hash (a fuzzy hashing function) to compare the page screenshots with the hash of the real pages to assess the visual similarity. & False Positive Rate - 0.5\%, \newline False Negative Rate - 0.05\%\\ \hline
    
    Hara et al.~\cite{PW:Hara2009}	& ImgSeek \newline (Wavelet Transform) & N/A & 2,262/\newline 521 & Extracts the image shown on the target web page, then uses ImgSeek to search for similar images in the database containing legitimate sites, phishing sites, and undeterminated sites, and checkes the output similarity to perform classification.  & Recall - 82.6\% \newline False Positive Rate - 18.0\%\\ \hline
    
\hline
\multicolumn{5}{c}{\textbf{Logo Feature Representation Approaches}} \\ \hline
\hline
      
   Wang et al.~\cite{PW:Wang2011} & SIFT & N/A & 219/230 & Captures a screenshot, then uses SIFT to match the screenshot against a logo database, and finally checks for keypress and the fact whether the DNS matches with the legitimate brand holder to decide if it is phish. & Recall - 90\% \newline False Positive Rate - 3\% \\ \hline

    Zhou et al.~\cite{PW:Zhou2014} & SURF and EMD	&100*100 & 2,129/\newline 1,372 & Combines and Compares logo detection and web page snapshots using SURF and homography transformation for possible detection of phishing websites. & Recall- 90\% \newline False Positive 1.46\% \\ \hline

    Afroz et al.~\cite{PW:Afroz2011} & SIFT, Fuzzy Hashing, and OCR	& N/A  & 1,000/\newline 200  & Uses SIFT to extract logo features form a web page, and performs image matching based on fuzzy hashing, OCR and SIFT algorithms.  & Recall- 90.2\% \newline False Positive - 0.5\% \\ \hline

    Fatt et al.~\cite{PW:Fatt2014}	& Google search-by-image API & N/A 	&500/500 & Extracts the favicon of website, then utilises Google search-by-image API to search for relative information, and then determines if the suspect page is a phishing page based on search results. & False Positive Rate - 5.4\%, \newline Recall - 97.2\%\\ \hline

\hline 
\multicolumn{5}{c}{\textbf{Hashing Approaches}} \\ \hline
\hline

    White et al.~\cite{PW:White2012} & Perceptual Hashing & N/A & - & Takes a page screenshot, then creates a hash of the image, and finally calculates the Hamming distance between the image hashes of the suspect pages and authentic pages. & \\ \hline
    
    Tian et al.~\cite{PW:Tian2018NeedleIA} & OCR and Image Hashing & N/A & 1,224/\newline 657,663 & Utilises OCR to extract texts from the screenshots of web pages as OCR features, and applies the Image hash (a fuzzy hashing function) to compare the page screenshots with the hash of the real pages to assess the visual similarity. & False Positive Rate - 0.5\%, \newline False Negative Rate - 0.05\%\\ \hline
    
    Fotiou et al.~\cite{PW:Fotiou2012} & Perceptual Hashing & N/A & 100/100 & Applies perceptual hashing on web page screenshots, then calculates the Hamming distance between the image hashes of the suspect pages and authentic pages. & False Positive Rate - 1\%, \newline Recall - 81\%\\ \hline

\hline 
\multicolumn{5}{c}{\textbf{Neural Network Approaches}} \\ \hline
\hline

    Borgolte et al.~\cite{PW:Borgolte2015} & Stacked Auto Encoder and MLP & screenshots(1600*900) and window(160*160)	&925,817/\newline 255,490	& Captures the snapshots of websites, extracts a specific window from each snapshot, and then uses a sliding window approach on the snapshot to perform defacement detection based on deep neural network.		& Recall - 97.42\% - 98.82\% \newline False Positive Rate \newline - 0.55\% - 1.53\%
    \\ \hline

    Adebowale et al.~\cite{PW:Adebowale2018} & SIFT, Classical Neural Network with Fuzzy Logic & N/A	& 6,843/\newline 6,157 & Extracts features of images, frames and text, and uses the hybrid features as input to a fuzzy neural network to classify the web page. & 	Recall - 98.3\% \newline False Positive Rate - 3\% \\ \hline
    
    Abdelnabi et al.~\cite{PW:abdelnabi2019} & Triplet CNN & screenshots(224*224) &1195/683 & Utilises triplet networks to train all screenshots with random sampling, then fine-tunes the weights of model by iterative training on hard examples, and finally (based on embeddings) performs classification by checking the distance with a threshold distance. & ROC Area-0.9879\newline Recall -95.81\% \newline False Positive Rate - 6.88\% \\ \hline

\end{tabular}
\end{table*}

\subsection{Summary of visual similarity based phishing detection:}

To summarise, in this section, we surveyed research work that used computer vision techniques to detect phishing web pages that are visually similar to target web pages. The motivation of using computer vision methods in this setting is the fact that many malicious websites increasingly use obfuscated codes using client-side scripting languages such as Java-scripts and it is relatively easy to create a website that shows no or minimum similarity to the original website when text features extracted from  HTML and Java-script codes are used, yet shows very high visual similarity to the user once the web page is rendered in a browser. This was evidenced by the results; many work reported that image-based methods achieve higher performance compared to text feature-based methods. 

We presented these work under three topics; \emph{\bfseries{i) image  feature-based  methods}}, \emph{\bfseries{ii) image hashing-based methods}}, and \emph{\bfseries{iii) neural  networks-based  approaches}}. \emph{Image feature-based approaches} create features such as SIFT, SURF, HOG, and CCH from the full screenshot of the web page, segments of the page, or embedded images/logos and measure how similar are those to the features from a target web page.  \emph{Image hashing-based methods} improve the speed of the similarity check by creating various hashes from images, and thus converting the time-consuming EMD and L2 distance calculations to more efficient Hamming Distance calculation. Finally, \emph{neural  networks-based approaches} leverage the recent advances in deep learning methods and eliminate the requirement of manually creating the features. We found more work in the category of image feature-based methods, but very recent trends in using deep learning-based methods. We summarise all the work we surveyed in Table~\ref{Tab:PhishSummary}. 

We next discuss several challenges and limitations we observed over all the work and worth addressing to make the research contributions more effective and can be deployed in industrial solutions.

\begin{itemize}
    
    \item \textbf{Datasets:} Almost all the work collected their own dataset using the PhishTank website which is a  crowd-sourced phishing verification system where users submit links to suspected phishing pages and other users ``vote'' if the reported page is a phishing page or not. It is relatively easy to build a sizeable and good quality dataset using PhishTank API that include the links to the phishing web page as well as the target. As a result, the majority of the work used their own dataset. A limited amount of work used data collected from their own institutions (i.e. by harvesting URLs from organizational emails) or crawled social media plot to circumvent the short lifetime issues of phishing web pages. For example, in PhishTank links, there is a delay between someone reporting the link and others voting for it as possible phishing. By the time this flagging happens, the phishing site might be already down, eliminating the possibility of obtaining a screenshot.


    
    \item \textbf{Performance Comparison:} As a consequence of different work using different datasets, it is not possible to quantitatively compare the performance methods. Analogous to the established datasets for various computer vision tasks such as MNIST~\cite{lecun1998gradient}, CIFAR 10 \& 100~\cite{krizhevsky2009learning}, ImageNet~\cite{deng2009imagenet}, and COCO~\cite{lin2014microsoft}, a common web page screenshot dataset would enable more re-producible research in this area. We noticed that, there are many new phishing website databases released between 2018 to 2019 such as Phish-IRIS dataset~\cite{phish-iris}, anti-phishing dataset by Universiti Malaysia Sarawak~\cite{malaysia}, and CIRCL dataset~\cite{falconieri2019} that have the potential to become standard datasets for phishing detection.
    
    
    
\item \textbf{Deep Neural Networks (DNNs):} We found that several recent work started using deep learning methods on web page screenshots to compute visual similarity, which is not surprising at all given the huge success of deep learning methods in computer vision. However, the current set of work use basic blocks of deep learning, such as MLPs, autoencoders, and CNNs. However, web page images are very diverse and rich in content, and the phishing tactics are becoming more and more advanced. As such more advanced deep learning methods for salient object detection and segmentation~\cite{li2016deep,liu2016dhsnet,li2016deepsaliency}, image comparison~\cite{hoffer2015deep,wan2014deep}, and neural image hashing~\cite{erin2015deep,lin2015deep,lai2015simultaneous} can be leveraged in this area to build more efficient and accurate phishing detection solutions.
     
    
    \end{itemize}

%% file: Section_04.tex
\section{Visual Similarity Based Malware Detection}
\label{sec:MDA}


More recently, computer vision methods help in alleviating the limitations of traditional malware detection systems such as detection of obfuscated malware or a computationally less expensive detection based on malware behaviour. Computer vision methods provide visual aids in detecting unknown malware to alert on anomalous behaviour patterns quickly. Visual representations of malware patterns also have the advantage of offering a summarised picture of possible attacks, thus speeding up the detection or classification of malware samples. For instance, if the frequency of API calls or usage of system resources is too high in an image representation, then a sample is classified as malware. Moreover, the fact that recent malware such as DuQu~\cite{duqu0} and Hammertoss~\cite{intelligence2015hammertoss} exploited the structure of image files (JPEG and PNG) either to communicate with the Command and Control (C\&C) server or to follow commands hidden under images files on a victim's machine. Finally, some malware display a familiar, trusted icon to users to trick them into clicking the malicious program, also motivates malware analysts to utilise computer vision techniques for possible malware detection. 

In this section, we survey the existing computer vision-based malware detection or classification approaches and categorise them based on their method of detection, according to the categorization we introduced in Section~\ref{sec:taxanomy}. We found a significant amount of work fitting into the categories of \textit{\textbf{i) image representation-based methods, ii) image feature-based methods, iii) image hashing-based methods}}, and \textit{\textbf{iv) neural network-based methods.}} 

\subsection{Image representation-based methods}
\label{sec4a}
Image representation-based methods visualise behavioural activities of malware as images and then compare them against the benign/normal activities of a system or a network. Malware behaviour refers to what the malware does, exhibits, or causes to its environment during live execution. Such behaviour aspects could be represented through monitoring changes to operating system resources during malware execution, capturing malware's API call sequence, malware's I/O request packets (IRP), or malware's network activity. It is possible to represent such behaviours as visual fingerprints (e.g. RGB colours, matrix, or intensity levels). For example, a visual fingerprint can be the frequency of calls to each API in RGB colours, where red denotes the most frequently called API (e.g. HTTP outgoing packet send request) and the green denotes the least called API (e.g. invoke anti-virus).  

Image representation-based methods for malware detection have several advantages. First, the image representation-based analysis helps in detecting obfuscated malware. For instance, a visual fingerprint of an obfuscated malware can be generated by executing it in a sandbox environment and then tracing the functionalities performed by malware. If the fingerprint matches with existing malware fingerprints, then it is considered malcious, otherwise benign. Similarly, it is possible to detect obfuscated malware, which has an icon disguised as a benign app, by using properties such as RGB pixels colours.
In this case, similarity analysis over the sets of images can provide insights into malware sample relationships. Second, the behavioural analysis of malware can help to detect new and unknown malware types. For example, if a new program (malware) communicates very frequently with C\&C server, then thread graph can be used to view a temporal behaviour of a new program, showing more often connections to a specific IP address. Similarly, behavioural to colour mapping, may also be used to identify potentially malicious programs by defining high-intensity colours (e.g. dark tones) to system-related activities such as system shut down, system files deletion, copy of system files, sending password files to a remote server etc. Finally, an image-based analysis may help analysts in highlighting current trends (e.g. icons, logos, hidden images, frequently used system functions, sequence/series of functions layout), in which malware authors are attempting to trick users into running malicious programs. 


One of the early work in visualizing and analyzing malware behaviour was by Trinius et al.~\cite{trinius2009visual}. The proposed work first applied an abstraction method to summarise the reports of the sandbox. Then classical visualization techniques, treemaps and thread graphs are used to visualise the percentages of API calls as well as malware behaviour. The abstraction method groups API calls performing similar functionalities (e.g. all API calls related to file system activity groups to one cluster). These API calls order arguments according to their significance in performing functionality.  The purpose of Treemapping is to help in viewing the distribution of operations, e.g. whether the main task lies in the area of network interaction or interaction with other processes. On the other hand, thread graphs help in visualizing the temporal behaviour of a sample.



To validate the effectiveness of this work, a large database of malware samples was collected using honeypots. Then the behaviour of these samples was analysed by executing them for two minutes in a controlled instrumented environment (sandbox).  During execution, sandbox records all the system-level activities (e.g. loaded system libraries, outgoing and incoming network connections, accessed or manipulated registry keys, and reports in XML format). To expand the analysis, the authors also executed the associated data files such as word or adobe acrobat files in a sandbox environment. After the execution, the abstraction method transformed XML reports to a short format; for example, by grouping together API calls having similar functionalities. Treemaps and thread graphs then helped in visualising the abstracted reports. The work claims to be effective against the classification of unknown samples to malware families by analysing over a set of 2,000 malware samples.

The advantage of this work is that it partly circumvents the problems of code obfuscation techniques such as packers or crypters. A packed malware is a modified form of a malware that is compressed using a compression program and is decompressed at runtime using a stub appended with a malware file. Therefore, when a packed malware executes in a sandbox environment, it unpacks itself and then it is possible to study its behaviour. Moreover, this method is capable of detecting malware hidden in data files such as adobe acrobat or Microsoft word files. However, the approach is not fully automated and require human intervention to compare thread graphs or treemaps of benign and malicious files. Another drawback of this approach is the fact that malware may detect the presence of the analysis environment (sandbox) and then behave differently.  This approach is not capable of detecting such behaviour.

Following the work of~\cite{trinius2009visual}, Han et al.~\cite{han2013malware} also proposed a novel method to detect and classify malware by constructing image matrices from the opcode sequence of executable files. The purpose of the image matrix is to detect malware features conveniently and to calculate similarities between different malware faster than other visualization methods. First, 
OllyDbg or IDA Pro dissembles binary files to extract binary information, and then opcode sequence is divided into blocks. These blocks are assigned (x,y) coordinates and RGB colours using \textit{Locality Sensitive Hash Function (SimHash)}. The RGB values are then plotted on their corresponding coordinates in a block of dimension $8 \times 8$ to get an image matrix. After forming a matrix for the whole image, ``selective area matching'' is performed using vector angular-based distance measure algorithm~\cite{androutsos1999novel} to calculate the similarities between image matrices of different images. The image matrices sizes are set to $256 \times 256$ pixels. The proposed method was applied to 10 different malware families, and the results were recorded. Results indicate that the image matrices from the same malware family have 0.95 similarity on average, but, those from different families have 0.325 similarity on average. Though this work shows promising results in classifying unknown samples to malware families, it was tested on a small number of malware samples. Also, it is unclear whether the authors collected malware samples or use some public dataset.


Similar to the above approaches,~\cite{MD:Han2014} also proposed a malware visual analysis method that generates RGB-coloured pixels on image matrices using the opcode sequences. However, to reduce the computational overheads, authors extracted the opcode sequences only from the blocks that are related to staple behaviours such as functions and API calls. The angular-based distance measurement algorithm was used to calculate the similarities between images matrices. 

In the paper~\cite{MD:Han2014}, authors performed two sets of experiments, static and dynamic analysis. In a static analysis experiment, 290 malware samples from 16 malware families, were tested. This experiment did not contain any packed or obfuscated malware samples as the purpose was to extract the basic blocks through static analysis using a disassembler.  Based on staple behaviour, the basic blocks helped in selecting major blocks, which were used to generate image matrices using opcode sequences. The image matrices were compared against each other. The dynamic analysis experiment consists of 560 malware samples from 14 malware families in which the packed and non-packed malware samples coexist. These samples generated the dynamic traces, and then basic blocks were extracted through repetition filtered technique. Afterwards, the major blocks relating to suspicious behaviours and functions were selected. The proposed method was then applied to these major blocks to generate image matrices and to analyze similarities. For both experiments, the sizes of the generated image matrices were $256 \times 256$ pixels. The static analysis shows an accuracy of 98.96\% whereas, dynamic analysis shows an accuracy of 97.32\%, receptively. The above two approaches~\cite{han2013malware, MD:Han2014} suffer from the drawback of disassembly of malware files, which make them ineffective for packers and encrypted malware. Hence, the characteristics of image representation based methods to analyze packed or encrypted malware, are not fully utilised by these methods. Authors also did not discuss any results or an extension of their work for packed and encrypted malware.

In another scheme, Han et. al~\cite{han2015malware} converted binary files into images and entropy graphs to detect and classify malware variants. The method first converts a binary file into a bitmap image using the ``Bitmap Image Converter''. It then calculates the entropy values of a greyscale bitmap image to generate the entropy graph. As the final step, the entropy graph similarity calculation algorithm detected unknown samples to a malware family. The method was tested on two sets of data collected from VX Heavens \cite{bontchevgood}. In the first dataset (consisted of 24 benign binaries and 27 malware binary files from 8 families) determined the threshold to minimise false positives. After setting the threshold to 0.75, the second dataset was tested with a total of 1,000 malware binary files from 50 families, and achieved an average accuracy of 97.9\%. The collected malware binary files were Backdoors, Trojans, Viruses, and Worms that are executable files in the Windows operating system.

This work has an advantage over previously proposed methods in terms of detection timing. It has less computational overhead than texture analysis methods, such as GIST~\cite{nataraj2011comparative}. The time to calculate the similarities of all the benign and malware binary files was ranged from 35.29ms to 1.39ms. However, the problem with this work is that it cannot be applied to packed malware because the entropy of packed malware usually are very high and cannot indicate any specific pattern. Moreover, all the three approaches from Han et al.~\cite{han2015malware, han2013malware, MD:Han2014} have a limitation of inspecting only a particular aspect of malware patterns and behaviours and did not focus on analyzing unknown processes, system calls and their relations.

Shaid et al.~\cite{shaid2014malware} showed another method to visualise and classify malware files. Authors first created a behaviour-to-colour mapping and then generated the behaviour-based image by converting each behaviour of malware to a colour. In order to obtain behavioural data, an API call monitoring utility executed the malware in real-time and captured all user-mode API calls made by a malware sample. API call monitoring utility captures the information such as accessed system files, network connections and protocols used, modified memory content of a remote process. The reason to opt for user-mode API calls is its stability as these calls are rarely changed as compared to the lower-level system call, which differs a lot between different versions of the same operating system. Besides being stable, capturing user-mode API is also crucial in speeding up the process of understanding the workings of malware since user-mode API reflects the exact nature of a particular malware. Afterwards, API calls are grouped and sorted according to their level of maliciousness. Once arranged, colours are assigned to each group such that hot colours represent malicious activity while cold colours represent benign activity. Once the API-to-colour map is ready, a behaviour image of $64 \times 4$ pixels size was generated from each malware sample. The experiment was run on 1,101 malware samples belonging to 12 different families and showed an accuracy of between 95.92\% and 98.92\%.

This work did not focus on detecting malware; instead, their only focus is to classify unknown malware samples to malware families. Moreover, the authors did not explain how they collected malware dataset. The performance of this method was also not discussed as we assume that executing malware on a Virtual Machine and then performing colour mapping might require more time than other detection techniques.

A very recent work~\cite{chapman2018sad} proposed SAD THUG (Structural Anomaly Detection for Transmissions of High-value information Using Graphics) to detect malware hidden under the structure of image files. The method consists of two main phases, a training phase, for building a formal model, and a classification phase, to check whether files correspond to that model. Each image file decomposes into the sequence of segments of different length. Most segments start with a header or byte sequence that indicates their type. Often, their payload starts with another header that is needed to interpret the segment’s payload correctly. This inner header has a significant impact on how the segment is interpreted; for instance, these headers represent the purpose of a segment or even whether the segment should be ignored completely by most decoders. Hence, segments with different inner headers are written and read for different purposes and should thus be assigned different subtypes.  Therefore, this method further defines the subtypes of these inner headers to detect malicious purposes hidden under the structure of JPEG or PNG files. Once subtypes are defined, classifiers are trained with benign files that contain the start of a file, end of the file, start of each segment, end of each segment, length of each segment, and subtype of each segment.

To classify the unknown file, the method used two parameters $\alpha$, and $\tau$, where $\tau$ determines the number of times a sequence of segment appears during the training phase and $\alpha$ indicates the reasonable length of each segment. For instance, $\tau$ of 1 means any sequence of segments that have been observed once in training phase are benign, and $\tau$ set to high values makes classifier more restrictive. During classification, first, the transition of segments from one to another is checked, if that segment has not been observed during the training phase or observed to follow the previous segment, then the file is considered to be anomalous. After passing this stage, the length of segments is checked if it is under a defined range. If all checks are passed, the file is considered to be benign. The approach used a total of 270,000 JPEG and 33,000 PNG files downloaded from Alexa top 25 popular websites and the third-parties found during the recursively crawling. For JPEG files, this approach achieved a mean true positive ratio of 99.24\% while 99.31\% for PNG files. On the other hand, the mean true negative ratio was 99.32\% and 98.88\% for JPEG and PNG, respectively.


While this work is interesting and has shown promising results, it can be further improved. First, the method currently supports only a limited number of image files (i.e. JPEG and PNG) and could be evaded by any other file types that it does not support. Second, its effectiveness is dependent on the training dataset. A training dataset that is not representative enough for all types of benign data observed in the classification phase may increase the false positive ratio.

To detect the repackaged apps and phishing malware, Sun et al.~\cite{sun2015droideagle} proposed the DroidEagle tool that compares the visual characteristics and components in the XML layouts of Android apps. The tool leverages the app repository analysis, named as RepoEagle, and host-based detection, named as HostEagle. RepoEagle performs large scale detection on app repositories, by extracting the layout tree as the fingerprint, which can indicate a visual structure of an app. The similarity between two layout trees is then measured using Layout Edit Distance (LED). LED measures the similarity between two layout trees by counting the minimum number of operations required to transform from one layout tree to another tree. The operations in layout tree transformation are node deletion, node insertion and node substitution. A small value of LED implies that the two layout trees are very similar, indicating that apps may be impersonated. On the other hand, HostEagle is a lightweight mobile app which conducts a real-time detection on mobile phones. This module extracts a layout tree and calculates the hash value of the layout tree as the visual signature. By comparing the signature of the app in the remote server, it can decide whether this app is a malware or not. Authors crawled and collected 100,096 apps from the Google market and various third-party markets. Results show that RepoEagle found 1,298 visually similar apps from a repository whereas, HostEagle is capable of determining whether a downloaded mobile app is a repackaged app or a phishing malware. 

Despite promising results, this method suffers from several limitations. First, all visually similar apps may not be necessarily similar in XML layouts, and it is necessary to consider the similarities in images. Second, app developers are starting to use code encryption methods, thus accessing codes and layout files may not always possible. Third, dependence on specific aspects related to one operating system will not allow making comparisons between heterogeneous app markets, and in such situations, only metadata and image similarity are meaningful.  






\subsection{Image feature-based methods}
\label{sec4:feat}

While image representation-based methods are capable of detecting obfuscated/encrypted malware and can also detect/classify malware and its variants through behavioural information, they are less efficient and robust for a large malware dataset~\cite{li2010challenges}. Consider the dynamic analysis of a malware binary, in which binary is executed, provided time to run its program logic, converted into an image based on call sequences, and finally the image is compared for possible malware detection. This whole process takes a few minutes (e.g. 3- 5 minutes) for a single binary sample. Even if the procedure is streamlined for 30 seconds, it will take years to process millions of malware samples such as Symantec Malware Corpus. In this regard, feature-based methods are highly accurate and efficient to detect or classify unknown sample, as compared to image-representation methods~\cite{ nataraj2011comparative}. 

In feature-based methods, a malware sample first converts into an image, and then image-related features are extracted. A given malware sample is generally read as a vector of 8 or 16-bit unsigned integers and then organised into a 2D array. The 2D array can then be visualised as a greyscale image in the range [0,255] (0: black, 255: white). After that, image features such as SIFT, SURF, and Gabor are extracted and fed into a detection system (either a similarity checker or a machine learning classifier) for possible classification. We evaluate each method under this category based on similar aspects mentioned in Section~\ref{sec:PWA}, \textit{i.e. input data type,  computer vision methods used, datasets, and performance}.

Nataraj et al.~\cite{MD:Nataraj2011} is one of the early work fitting into this category. The proposed method assumed that greyscale image of the same malware family has similar layout and texture. Authors extracted 320 GIST features and used k-nearest neighbours with Euclidean distance on top of those features to classify unknown malware. Authors evaluated the performance of the solution using malware executables submitted to Anubis analysis system~\cite{anubis} and showed an average accuracy of 98\%  over a database of 9,458 samples and 25 malware families.

One advantage of this work is that it can detect packed malware as it is based on the assumption that the images of the same malware family when packed with the same packer, appear similar. Authors achieved a classification accuracy of 98.08\% when they packed malware of the same family and applied the proposed method on them. Authors also compared their performance with existing methods based on bi-gram executions. They extracted bi-grams distributions from the raw dataset and obtained a classification accuracy of 98\%, which is same as their approach. However, the time taken to extract bi-grams are 5 seconds, and the time to classify a malware sample is 56 seconds. In contrast, the proposed approach computed GIST features in 54 milliseconds and classified a sample in 1.4 seconds. The high efficiency of the proposed method is partly explained by the fact that the feature vector length used to characterise a malware image is about 320 whereas the distribution based analysis using the bi-grams needed about 65K elements. However, a major limitation of this approach is that visual representation only considers images embedded in a resource section of an executable. This limitation allows an attacker to easily bypass detection either by relocating sections in a binary or creating less obvious patterns.




In a later extension~\cite{nataraj2011comparative}, Nataraj et al. compared the performance of the initial work~\cite{MD:Nataraj2011}, named binary-texture analysis, with contemporary dynamic techniques. While the classification accuracy of the binary-texture analysis is comparable with dynamic analysis methods, however, it is 4,000 times faster than dynamic techniques. In addition, the texture-based approach is resilient against packed malware samples. They confirmed that the contemporary packing strategies perform a monotonic transformation which fails to conceal common structures present in the malware binaries, thus easily detectable. Authors used two dynamic analysis techniques. The first one is a commonly used approach called system-call level monitoring which generates a report of all system calls and their arguments performed by malware binary. The second approach is a forensic snapshot which compares features before and after a malware infection.

Authors conducted three different experiments; the first experiment was conducted on 393 malware binaries and showed the classification accuracy of 98\% and 95\% for dynamic and binary texture analysis, respectively. In the second set of experiments, Malheur dataset~\cite{malheur} was used for comparison that comprises of Reference dataset with 3,131 malware binaries from 24 unique malware families, and an application dataset of roughly 33K binaries that range from malicious, to unknown, to known benign. This experiment showed the classification accuracy of binary texture analysis ranging from 86\% to 97\%. In the third experiment, VX heavens dataset was used that consists of over 63K malicious binaries from 531 unique families. The binary-texture analysis, in this case, achieved 72\% accuracy. The reason for this high accuracy is that packed malware variants that belong to the same family exhibit visual similarity with one another, which can be detectable through binary-texture analysis method.  Another reason is that a single malware family is packed with one or two packing software and not more. Finally, there is a fraction of the malware that is unpacked. Thus, all the reasons mentioned above aid binary-texture analysis in detecting malware with high accuracy. However, the limitation is that this method is vulnerable to knowledgeable adversaries who can obfuscate their malicious code to defeat texture analysis.

Kancherla et al.~\cite{kancherla2013image} proposed a malware detection method that first converts a binary executable file to an 8-bit 1-dimensional vector and then the 8-bit value converts to the intensity of the pixel. Next, this single-dimensional vector converts into a 2-dimensional vector, known as \textit{byplot}. This byplot (image) has a fixed width based on the size of the file. From this image, the authors extracted 534 features that included 512 Gabor-based, 16 wavelet-based, and six intensity-based features. The purpose of using wavelet and intensity-based features is to examine if light-weight features offer comparable classification accuracy against computationally expensive Gabor-based features. Authors performed three different sets of experiments using only Gabor features, using only intensity and wavelets, and using all features. Results showed that light-weight intensity and Wavelet-based features produced slightly better classification accuracy as compared to Gabor based features. The resulting accuracy was around 94.32\% for the light-weight features, whereas Gabor features showed 93.23\% accuracy. These experiments were conducted on offensive computing database containing 25,000 malware samples and from other sources like source forge to collect 12,000 benign samples.


This work is an improvement to~\cite{MD:Nataraj2011} as it shows that light-weight features such as intensity-based features and wavelet-based features offer better classification performance when compared to computationally expensive Gabor based features.  Few other works have very similar approaches~\cite{makandar2015malware,makandar2017malware,makandar2018trojan} with the only difference being the final classifier and the dataset on which the solution is evaluated.

Ahmadi et al.~\cite{ahmadi2016novel} emphasised on the convenience to select simple features to design light and efficient malware-detection system. Instead of extracting features directly from a Portable Executable (PE) file, authors calculated a set of features from different categories in hex and assembly view of PE. The feature categories in hex view are n-gram of bytes, the byte-level entropy vector, metadata (e.g. size of the file, address of the first bytes sequence), the histogram of the length of strings, image-features (Haralick features, or the Local Binary Patterns features) whereas, the categories in assembly view are frequency of some specific operation codes, frequency of API calls, frequency of use of the registers, and characteristics of sections in a file. With each category, there is a range of features; for instance, in image-based features, there are a total of 160 features. In order to combine feature categories such that accuracy in classifying/detecting malware increases (i.e. best subset of relevant features), forward step-wise feature selection was used. In forward subset selection, initially, a model contains no features and then gradually augments the feature set by adding more features to the model, one by one. In order to check the performance of each feature category and combination of categories, XGBoost classifier added features from each category into the model one by one. Authors used Microsoft released a dataset containing  21,741 malware samples, where 10,868 samples are used for training, and the rest are used for testing. Results showed that image-based category has a classification accuracy of 95.5\% however, with the combination of other features, this accuracy reaches to 99.40\%. 






The subset of features utilised in~\cite{ahmadi2016novel} is still considered to be expensive and have chances of reducing the efficiency of a malware detection system. Therefore, Zhao et al. ~\cite{xiaolin2018research} proposed an approach that used malicious code homology analysis method (based on texture fingerprint clustering) for possible detection of malware. The approach first extract features from a binary program, create texture fingerprint from the features and then use a clustering algorithm to group the texture fingerprint of malware into families. Finally, the homology analysis matched the unknown sample with one of the templates of malware families stored in the database. To extract features, B2M algorithm converted unknown samples to an image, and then an improved SIFT algorithm was applied to create a multi-scale space. Once space is created, the feature descriptors of a sample is extracted and fed into DBSCAN for clustering into family libraries. The homology analysis is the same as image recognition, where an unknown sample is compared with the family library to detect its maliciousness and classify into a specific library. The method was tested on 1,789 samples from 11 different malware families, collected from Bigdata Innovators Gathering (BIG) 2015~\cite{bigdata2015}. Results show the average accuracy of 77.6\% against different malware families. 

A recent approach by Masouleh et al.~\cite{silva2018improving} classifies windows malware (PE files) using icon-based features. The method used three types of features generated from the original icon pixels: i) Summary statistics features, ii) Histogram of oriented gradient (HOG) features, and iii) Autoencoder (AE) features. Summary statistics features are the mean and standard deviation computed for all the pixels and across all three channels, on each channel separately, and by splitting the icon image into nine different sections. HOG features are captured by sliding a small window over the image and then computing the gradient of the image within that window. To extract AE features, neural-network is trained with a large set of icons. In total, 1,114 features were extracted. Next, hierarchical density-based spatial clustering (HDBSCAN) and K-Nearest Neighbor algorithm were used to cluster the icons with the purpose to reduce the data dimensionality. In order to test the efficacy of the method, a balanced sample of publicly available PE files was obtained from VirusTotal with 1,138 benign and 1,138 malware files. The dataset is split into 80\% testing and 20\% training and then evaluated using Logistic Regression (L1, L2) and SVM classifiers. Logistic Regression (L1) shows better performance than the other two classifiers with an accuracy of 84.4\%. 

This work~\cite{silva2018improving} may also fit into the category of neural network-based methods. Still, we have included it here because the neural networks do not play a major role in the methodology to detect malware. The neural network-based classifiers were primarily used to validate the efficacy of the proposed methodology. This work also suffers from several limitations. First, it is not validated on a large number of malware PE samples, and therefore its efficiency cannot be properly determined. Second, extraction of 1,114 features is quite a cumbersome task and time-consuming. Authors did not show the performance of the method with respect to time. Third, the technique is not compared against previously proposed icon-based malware detection technique~\cite{andow2016study}. Finally, the classification accuracy is quite low as compared to other techniques that were tested on Microsoft Challenge dataset. For instance, Ahmadi et al.~\cite{ahmadi2016novel} showed 95.5\% of classification accuracy which is 10\% higher than ~\cite{silva2018improving}. Furthermore, approaches in neural network-based methods also used Microsoft dataset and achieved accuracy ranging from 95\% to 99.8\%~\cite{kim2018detecting, kim2018zero, yan2018detecting, le2018deep, kalash2018malware, cao2018efficient }.


In~\cite{MD:Kywe2015}, authors proposed a text-based and content-based image retrieval method to detect camouflaged apps on mobile market places. For each app, four external features are extracted; namely app name, description, icon and screenshot. App name and descriptions are handled by text retrieval systems, while app icon and screenshots are handled by image retrieval system. Authors used unofficial Google API to crawl app info, such as id, name, developer, rating as well as application description, icon and screenshots from the official Android market. Total of 30,625 applications was crawled for the experiment. After crawling, each app was assigned four indexes which are name index, description index, icon index and screenshot index. For the image indexing, colour correlogram algorithm was used. The last step is to detect camouflaged apps by querying the indexed database. For each queried app, the method retrieved apps which have similar user interfaces but are from different developers. Both text-based and content-based systems are used to calculate the cosine similarity scores between the query app and a set of indexed apps. Out of 30,625, this method identified 477 camouflaged apps (approx. 1.56\%); however, the manual review showed that 44 apps were falsy flagged camouflaged with 9.22\% of the false positive rate.

\subsection{Image hashing-based methods}

The work discussed in above Section~\ref{SubSec:ImageFeature} showed high accuracy for the known malware families; however, the feature extraction and matching (similarity calculation) process still requires a time investment per malware that does not scale well to the high volume of malware samples at run-time. Consider the extraction and similarity calculation of 512 Gabor-based features~\cite{kancherla2013image}, 1,114 AE features~\cite{silva2018improving}, or even 160 Haralick features~\cite{ahmadi2016novel} that require sufficient time to match with the known malware stored in a database, thus highlighting the need of an efficient feature matching algorithms, as highlighted by~\cite{xiaofang2014malware}. A body of work addresses this problem by proposing \textit{image-hashing based methods}, which we discuss next in this subsection.


To improve the efficiency of malware classification, Xiaofang et al.~\cite{xiaofang2014malware} proposed a novel distance (similarity) metrics based on the locality-sensitive hashing (LSH) schemes.  The approach classifies malware by calculating the distance metrics on the fingerprints of malware content. First, a binary image is obtained from a malware sample, and then a feature fingerprint of 64*64  dimension is computed using SURF algorithm. Afterwards, a fast fingerprint matching is performed with the LSH to get the topmost visually (structurally) similar variants. Here a euclidean distance is used as a metric to get a matched malware set. The proposed scheme is applied to 25 malware families on a total of 8,410 malware samples. The precision, recall and F-Score is around 80\% to 90\% for varying dataset sizes. Though this work claims to improve the efficiency of computer-vision based malware detection systems by using LSH scheme, the authors did not provide any results of performance comparison in the paper to prove the claim. 


Another body of work used image hashing-based methods to detect mobile malware and greyware. Malisa et al.~\cite{malisa2016mobile} detected two types of mobile impersonation attacks; \textit{repackaging} and \textit{phishing}.  Authors proposed visual impersonation detection system that runs mobile app inside an emulator, explores its user interface dynamically, and extracts app screenshots using GUI crawling techniques. The extracted screenshots are then compared with other apps to detect the case of possible impersonation. If the comparison value is above a threshold (i.e. near match or exact match), apps are labelled as impersonated apps. The Local Sensitive Hashing (LSH) technique is used to check the similarity between screenshots. Unlike other hashes, LSH maps similar data to the same hash code with high probability, maximizing the probability of a collision for similar inputs. The method was deployed on the Google Cloud Compute platform and tested on 150,000 mobile apps with 4.3 million screenshots. Results indicate the detection of 43,904 impersonating apps with 15\% FPR. 

The proposed system is capable of detecting impersonated apps that are using evasion techniques such as obfuscation, as the detection purely relies on the visual appearance of the examined applications. Another advantage of this system is that it can efficiently analyze a large number of Android apps before their deployment at Play Store. Moreover, the detection system neither requires human input nor application-specific knowledge as it supports the automated analysis of apps at a larger scale. Nonetheless, this method inherits all the resource limitations of dynamic analysis as each app runs in the emulator.


Besides mobile malware, greyware is another emerging threat to mobile apps. For instance, ad fraud apps could be classified as greyware because of the reason that such apps contain annoying, undesirable or undisclosed behaviours that cannot be classified as malware. In this regard, Andow et al.~\cite{andow2016study} first identified nine categories of greyware based on installation and runtime behaviour. Later, authors developed lightweight heuristics for triage of the seven categories of greyware on Google Play that leverages text analytics and static program analysis. Among all the greyware categories, imposter apps are detected using computer vision methods. To detect imposters, first, the similarity of apps’ titles and developers’ names are computed and then \emph{fuzzy hashing} is used to score the similarity of app icons.

Authors retrieved a list of APKs and app metadata such as user reviews, description, and developer names from the PlayDrone Archive~\cite{playdrone} and separate them in four different datasets. The dataset to detect imposters contained 50 most popular apps for each app category in both “free” and “paid” subcategories, which a total of 2,500 apps. Results indicate that manual review of 2,500 apps is reduced to 977 with exact title matches, of which 22 app pairs have similar icons. Though fuzzy hashing is suitable for the detection of slightly modified visual images; however, the efficient detection of impersonated apps requires techniques other than hashing (e.g. feature comparisons or deep learning classification methods). Moreover, the proposed imposter detection method was tested only on a small portion of apps, i.e.  2,500; this number is quite low as compared to millions of apps available at PlayStore.  

Alike \cite{silva2018improving,andow2016study}, Jiao et al.~\cite{MD:Jiao2015} also proposed a method called ImageStruct that detects repackaged apps using similarity measurement. The method extracted features from app images using pHash and compared the hash codes to detect similar apps. When the similarity of two apps with different signatures exceeds a certain threshold, one of the apps was considered as repackaged. Similarly, Long et al.~\cite{MD:Long2014} visualised similarity between malware samples based on their embedded graphical assets. First, images were extracted from malware using \textit{wrestool} ~\cite{wrestool} and converted to ,1024 dimensional vector. Then, the vectors were stored using Fast Library for Approximate Nearest Neighbor (FLANN) index. Finally, the top $n$ visually similar images were retrieved by the random tree method using $L2$ distances. The authors claimed that the method has a recall of 83\% and precision of 85\%. Since the method uses average hashing and FLANN indexing, it is fast; however, the performance is highly depending on the threshold values.


\subsection{Neural network-based methods}
\label{sub:mal_neural}


Similar to the motivation of moving into neural networks as mentioned in Section~\ref{sec:PWA}, in malware detection also multiple works proposed to use neural networks, especially deep learning based networks to eliminate the need of manual feature curation. In image-based neural network approaches for malware detection, the malware binaries are converted into greyscale image representation, and deep learning architectures are employed to learn the complex features (image patterns). Most commonly employed deep learning architectures are convolutional neural networks and LSTM (long short-term memory) recurrent neural networks.

Zhang et al.~\cite{zhang2016irmd} proposed a CNN to compare opcode images of a target with the opcode images generated by known malware sample codes to detect binary malware variants. First, the binary executable is disassembled, and an opcode profile for the binary is built by representing it as a list of opcode sequences of length two and the corresponding frequency. This sequence is converted to an image where each pixel value represents the information gain of the two corresponding opcodes. This image is further enhanced using methods in the likes of histogram normalization, erosion, and dilation. These images are then used to train a simple CNN with two sets of convolution and pooling layers followed by one fully connected layer.

Authors evaluated the classifier using a dataset of 9,168 malware samples belonging to 10 families and 8,640 benign samples. The results showed that image-based malware variant detector could achieve over 90\% true positive rate and true negative rate outperforms traditional machine learning-based classifiers such as naive Bayes and $k$ nearest neighbours especially when there is less number of training samples.

Kim et al~\cite{kim2017malware, kim2018zero, kim2018detecting} conducted a series of work on using generative adversarial networks (GANs) for zero-day malware detection. The idea stems from the GANs in computer vision that has found multiple applications in image generation, denoising, and resolution improvements. The key idea in related to zero-day detection is to use the \emph{generator function} of the GAN to generate new malware samples and use the discriminator function to predict whether the given sample is a malware or not. The intuition behind this process is that as the generator builds new malware representations and the discriminator has been trained to identify them, when a new malware that was not part of the training set emerges, the discriminator will be able to detect it.

In this regard, the first work by Kim et al.~\cite{kim2017malware} proposed transferred generative adversarial network (tGAN) that uses a pre-trained auto-encoder as the generator and a conventional CNN as the discriminator. The auto-encoder learns the characteristics of data using transfer learning method and generates malware images similar to real malware images. On the contrary, the CNN based discriminator tries to distinguish between real and fake malware images.  However, this approach suffers from the problem of instability in converging generator and discriminator to Nash equilibrium during the training process. Therefore, Kim et al. extended the idea and proposed two different approaches. In the first approach, Kim et al.~\cite{kim2018zero} proposed transferred deep-convolutional generative adversarial network (tDCGAN) that uses a deep auto-encoder (DAE) as a generator and deep convolutional GAN (DCGAN) as a discriminator. The advantage of using DCGAN is that it produces higher-quality images than original GAN, and the training system is relatively stable. Akin to above method, Kim et al.~\cite{kim2018detecting} also proposed another version of GANS called latent semantic controlling generative adversarial networks (LSC-GAN), that generates malware with arbitrary modified features and help detector to learn features with unseen observations.

In all the three approaches~\cite{kim2017malware, kim2018zero, kim2018detecting}, authors used a dataset from Kaggle Microsoft Malware Classification Challenge~\cite{bigdata2015}, which contains nine malware families. The accuracy of detecting malware in~\cite{kim2017malware} is 96.39\% for 90:10 ratio of training-testing data. For~\cite{kim2018zero} and~\cite{kim2018detecting}, authors achieve the average accuracy of 95.74\% and 96.97\% respectively. Nevertheless, the major problems in these approaches are the interpretability of the results obtained and the black-box nature of these systems. The difficulties with interpreting and understanding neural networks such as GANs, places a constraint on how these models can be modified. Also, the process of training large scale GANs is not a simple and trivial task. Designers of malware detection systems need an efficient GAN-based security system to defeat the rapidly evolving malware adversarial programs~\cite{yinka2019review}.

Ni et al. \cite{ni2018malware} proposed an algorithm named, Malware Classification using SimHash and CNN (MCSC), that converts SimHash values to greyscale images and then uses CNN to identify malware families. The algorithm combines multiple methods such as cascade hashing, major block selection, and bilinear interpolation to improve accuracy. The algorithm is divided into three parts: the first part extracts opcode sequence-based features and calculates the similarity between sequences using SimHash method. In this step, multiple cascading hash functions strengthen SimHash. Moreover, major block selection chooses opcode sequences that contain representative/informative feature. For example, the opcode sequences containing CALL instructions are considered as a major block because this instruction invokes APIs, library functions etc. In a second step, the SimHash value and bilinear interpolation convert opcode sequences to malware images. Finally, these images train CNN model with the aim to identify their families. Similar to other methods, this algorithm also used Malware Classification Challenge dataset containing 10,805 samples. The classification accuracy is 98.86\% on average, which is higher than other algorithms. However, this work can identify only known malware, and no solution was indicated that shows the integration of characteristics of new malware in the detection system. 

Akin to above, Yan et al.~\cite{yan2018detecting} also proposed a method, MalNet, that uses greyscale images and opcode sequences to detect and classify malware. The method uses CNN for extracting image-based features while Long Short Term Memory (LSTM) is used to learn opcode sequences features. Moreover, in order to overcome the gradient vanishing problem of LST, the truncated backpropagation algorithm is used along with a subsequence selection method to filter out benign opcode sequences. The processing of the MalNet is divided into two stages. The first stage is to preprocess malware sample data; it takes a binary form of a Windows executable file, generates a greyscale image from it, and extracts opcode sequence and metadata features using a disassembler tool. CNN and LTSM then process the greyscale image, and the results are sent to the stacking ensemble to integrate two networks' output and metadata. The experiments were conducted on 21,736 malware samples from Microsoft and 20,650 benign samples. The MalNet achieves detection accuracy of 99.88\% and TPR of 99.1\% with a false positive rate of 0.1\% respectively. In addition, the malware classification accuracy of nine families reaches up to 99.36\%.

To address the problem of data imbalance, Cui et al.~\cite{cui2018detection} developed a method that combines the CNN with the Bat algorithm. The Bat algorithm is used to balance the image dataset, and CNN is used to identify and classify the images. In this work, first, samples are converted into greyscale images and then classified using CNN. The size of an image determines the structure (layers) of CNN, whereas, the width of the image could be fixed based on the pre-defined standard given in the literature. For instance, an image file of ~100kb to 200kb size should have a fixed image width of 384. The method uses two different types of layers, namely convolution and sub-sampling layer. The former reduces the number of image parameters and preserves the main features such as translation, rotation, or scale invariance; while the latter reduces the dimensions of features, thus weakening the influence of image deformation. Both layer types improve the accuracy of models and avoid overfitting. The malware dataset was taken from Vision Research Lab consisting of 9342 greyscale images of 25 malware families. The achieved average accuracy is 94.5\%.

Yue et al.~\cite{MD:Yue2017} also addresses the problem of class imbalance by proposing a method that incorporates CNN and weighted softmax loss method for effective classification. Authors used vgg-verydeep-19 model with 43 layers and also placed two dropout layers between the three fully connected layers to prevent overfitting. The weighted softmax loss layer is then used as the last layer.  The method achieved 98.63\% accuracy with the Malimg dataset containing samples from 25 malware families which were highly imbalanced. The problem with this work is that it requires more computation as there is a total of 60 layers which are used for malware classification.

Le et al.~\cite{le2018deep} proposed a generic image scaling approach, where a raw malware byte code is interpreted as a one-dimensional image and is scaled to a fixed target size. The raw static byte code inputs to a convolutional neural network followed by a recurrent neural network. The approach uses three different models of CNN; i) conventional CNN, ii) CNN-UniLSTM and iii) CNN-BiLSTM. In the second and third model, a recurrent neural network (RNN) is used to improve the performance of malware classification. Similar to previous works, the authors used the Microsoft Malware Classification Challenge dataset containing 10,868 samples. The results showed that the CNN-BiLSTM has the highest accuracy of 98.20\% as compared to the other two models.

This approach is simpler than converting a malware binary file to a 2D image since one doesn't have to make a decision about the height and the width of the image. The image size is fixed to 10000 bytes. However, this work was intended to identify the malware class from nine classes and not to decide if a file is benign or malicious. Akin to above, Kalash et al.~\cite{kalash2018malware} also converted malware binaries to images and trained CNN. However, authors used VGG network architecture which is a variant of CNN.  The method was tested on two large datasets, Malimg and Microsoft malware, and achieved an accuracy of 98.52\% and 99.97\% respectively.

Jiawei et al.~\cite{su2018lightweight} also used CNN to detect and classify malware residing in IoT devices. The proposed approach uses a rescaling method to standardise the image sizes of malware samples to 64*64 pixels. Such images then serve as input to a CNN that determines whether the analyzed program is malicious or benign. To evaluate the performance, authors used malware captured by the IoTPOT honeypot \cite{pa2016iotpot}. The approach was run on 500 malware samples and achieved an average accuracy of 94.0\% for two-class classification and 81.8\% for three-class classification. (i.e., benign, Mirai, or Gafgyt). Analysis results showed that malware samples have distinct byte patterns in the padded areas and unused data sections. However, the authors admit that their approach is susceptible to complex code obfuscation. In addition, the padded areas and unused data sections can be easily modified by malware developers without affecting the behaviours of malware. 


Cao et al.~\cite{cao2018efficient} developed a malware identification and classification system based on CNNs. Firstly, the binary files of potentially malicious codes are converted into greyscale images, and then the images are fed into the CNN for malware detection. The proposed system works in an online environment where a browser plug-in checks if a file is malicious or not. If considered malicious, it is sent to the detection server where the CNN performs further analysis, and finally, the database is updated.  In another scenario, a user can also upload the malicious file on a website. The method used two popular malware datasets, Vision Research Lab and Microsoft Malware Classification Challenge with the 20 well-known malicious code families. The system achieved an accuracy of 95\%. 

Similar to above, Akrash et al.~\cite{akarsh2019detailed} classified malware families by converting malware binaries to greyscale images and using a hybrid CNN and Long Short Term Memory (CNN-LSTM). The hybrid model enables the extraction of temporal and spatial features which are later used to identify the malware family. To select a generalised model that is suitable to deploy at run-time, the authors conducted several experiments on Malimg dataset that contains 9,389 malware samples of 25 different families. The accuracy of the method ranges from 95\% to 97\%. However, this research work did not explain how it is different from other neural-network-based approaches.

A recent approach by Venkatraman et al.~\cite{venkatraman2019hybrid} used an image-based technique with deep learning architectures to detect and classify obfuscated malware into their respective families. The proposed approach performs an image comparison of different malware families as well as benign datasets to visually demonstrate the significant difference in the behaviour patterns of the malware families. This approach is different from other methods as it uses a self-learning system which is capable of detecting not only the known malware and the variants of known malware but also unknown malware. The architecture composed of three different subsystems, in which one subsystem uses unsupervised learning model and the other two use supervised learning models. First, pre-processing on samples are performed to detect packed and unpacked binaries. After this, image analysis extracts image related features from a pre-trained deep CNN model and then clusters in the image feature space using k-means clustering algorithm. The features are then fed into CNN. The approach was tested on 52,000 malware samples collected privately in previous work as well as commonly used public datasets for benchmarking. Results indicated accuracy ranges from 94.8\% to 96.3\% on average, whereas F1 score ranged from  90.3\% to 91.6\%. 

A method proposed by Darshan et al.~\cite{sl2019windows} also used a CNN to detect windows-based malware. However, their process to generate an image is quite different from the rest of the methods. It uses the run-time behaviour features (n-grams) of the PE files, selected by a feature selection technique to create images, and the generated images are then fed into the CNN for the detection. In the training phase, behaviour-based feature extractor records the behavioural features (CAT-API features) of the PE file, which is under execution in a Cuckoo sandbox. These features are then processed to derive n-grams. The n-grams are then passed through feature selection techniques such as Chi-Square, Mutual Information, Information Gain, and Relief to choose the best n-grams. The chosen n-grams are passed through the image generator that checks the occurrence of the n-gram present in the final feature set to construct an image for the n-gram file. Based on the presence or absence of the n-gram, the value 255 or 0 is written onto the image (0 represents that the n-gram is absent, and 255 indicates that the n-gram is present). Finally, each input image proceeded through two convolution layers, two sub-sampling layers, and two fully connected layers for CNN training. In the testing phase, the test image was created from the unknown PE files, which was then sent to the trained CNN to measure the detection ability of the trained CNN. Authors used 400 self-generated samples (200 benign, 200 malware) and a Malheur dataset~\cite{rieck2011automatic} to demonstrate the effectiveness of the proposed CNN-based Windows malware detector. Malheur dataset contains 3,282 benign and 4,151 malware files that included ten different types of malware. A set of experiments were conducted to demonstrate the classification ability of the proposed approach and compared with the chosen six machine learning-based classifiers. The method achieved maximum detection accuracy of 97.968\% outperforming machine learning-based classifiers.

To overcome the drawbacks of code obfuscation and high computational costs of analysing malware, Xue et al.~\cite{xue2019malware} recently proposed Malscore, which is based on the probability scoring and machine learning. The purpose of probability scoring is to decide if static analysis needs to concatenate with dynamic analysis.  First, greyscale images are generated from raw malware to train a CNN-based classifier. The output from the CNN classifier judges the credibility of each malware sample by comparing the score with the probability threshold. The samples with lower credibility are dynamically analysed by executing the malware in the sandbox and extracting native API call sequences. These API call sequences train n-grams and machine learning-based classifier. To understand this whole process, consider a packed/obfuscated malware that gets lower credibility from the CNN classifier. This sample will then be filtered out for dynamic analysis. In this way, this work improves the robustness for packed/obfuscated malware and also improves the classification accuracy. Similarly, to reduce the cost of detection, probability scoring in this work filters out most malware that get reliable classification results. The experiments were conducted on the dataset from VX Heaven website~\cite{bontchevgood} and after pre-processing, 174,607 malware samples from 63 malware families were used. The result showed that Malscore achieved 98.82\% accuracy while reducing the pre-processing and test time by 59.58\% and 61.70\%, as compared to previous methods.

Though highly efficient, this work did not use dynamic analysis as the primary detection method, but as a complementary method for static analysis to increase the robustness in detecting packed malware. Also, this work did not take into account the concept drift as it did not provide any incremental learning strategies to fortify re-learning ability. 

To detect counterfeit mobile apps, Rajasegaran et al.~\cite{rajasegaran2019multi} proposed a neural embedding-based approach that combines content and style embeddings generated from pre-trained CNNs. The method relies on the idea that a combination of style and content embeddings achieves better results in detecting visually similar app icons.  First, the app icons are fed into a pre-trained VGGNet for the extraction of content and style embeddings. Authors crawled and collected approximately 1.2 million apps from Google Play Store and showed that a combination of content and style embeddings achieved 8\% to 12\% higher precision and 14\% to 18\% higher recall, respectively compared to conventional methods. The precision and recall are further improved to 3\% to 5\% and 6\% to 7\% by adding text embeddings. Authors identified 7,246 potential counterfeits to the top-10,000 apps, out of which 2,040 may contain malware. Results further indicate that 1,565 of apps ask for at least five dangerous permissions and 1,407 have at least five extra third-party ad libraries.  

 

\begin{table*}[!h]
\tiny
\caption{Summary of Visual Similarity Based Malware Detection Approaches} 
\label{tab2:Mal_summary} \vspace{-3mm}
\begin{tabular}{p{1.8cm}|p{2.2cm}| p{2cm}|p{1.3cm}|p{5.5cm}|p{2.9cm}}

 \hline
\textbf{Research Work}	& \textbf{Computer Vision \newline Methods}	& \textbf{Input Image \newline Structure} & \textbf{Dataset Size/Family} & \textbf{Summary} & \textbf{Performance Metrics	}\\ \hline

\hline
\multicolumn{5}{c}{\textbf{Image Representation-based Methods}} \\ \hline
\hline

Trinus et al.~\cite{trinius2009visual} & Treemaps, ThreadGraphs, CWSandbox & (Generic) Size similar to original image & 2,000/13 & Applies an abstraction method to make malware reports easily understandable and to enable the extraction of unique features for malware detection. & Not Given \\
\hline

Han et al.~\cite{han2013malware} & SimHash, Vector Angular-based Distance Measure & $256\times256$ pixels of image matrices &  -/3 & Generates image matrices with (x,y) coordinates and RGB colours and then uses vector angular-based distance measure algorithm to calculate the similarities between image matrices. & Similarity (across same family) - 0.95 \newline Similarity (across diff. family) - 0.32 \\
\hline

Han et al.~\cite{MD:Han2014}& SimHash, Vector Angular-based Distance Measure & $256\times 256$ image & Static- 290/16 \newline Dynamic- 560/14 & Extracts the opcode sequences related to staple behaviours and uses angular-based distance measure to calculate the similarities. & Accuracy (Static)- 98.96\% \newline Accuracy (Dynamic)- 97.32\% \\ 
\hline

Han et al.~\cite{han2015malware} & Bitmap Image Convertor, Entropy Graph Generator, Similarity Calculation Algorithm  & (Generic) Pixels of original file size &  1,000/50 & Calculates entropy values of a greyscale bitmap image to generate the entropy graph and further measures similarity of entropy graph for the malware detection. & Accuracy (ROC) - 97.9\%\\  
\hline

Shaid et al.~\cite{shaid2014malware} & API-to-Colour Mapping, API Monitoring Utility, Similarity Measure & $64\times4$ pixel & 1,101/12 & Utilises API call monitoring utility to obtain behavioural data of an image by mapping to colour, and then groups malicious behaviour. & Accuracy - 95.92\% to 98.92\% \\
\hline 

Chapman et al.~\cite{chapman2018sad} & File Structure Decomposition & JPEG and PNG files of variable length & 270k JPEG and 33k PNG files & Decomposes each image file into sequence of segments, feeds into the training algorithm to build a formal model, and then classifies unknown samples. & TPR - 99.24\% (JPEG), 99.31\% (PNG) \newline TNR- 99.32\% (JPEG), 98.88\% (PNG) \\ \hline

Sun et al.~\cite{sun2015droideagle} & Layout Edit Distance (LED) & Apps XML Layouts in tree forms & 1,00,096 apps & Detects the repackaged and phishing malware in apps by calculating the LED between XML layout of android apps. & Visually Similar Apps: 1298 \\ 
 
\hline
\multicolumn{5}{c}{\textbf{Image Feature-based Methods}} \\ \hline
\hline

Nataraj et al.~\cite{MD:Nataraj2011} &	k-NN with Euclidean Distance & 320 GIST descriptors	& 9,458/25 & Converts binary file to a greyscale image and extracts 320 GIST features to classify malware using k-NN.  &  Accuracy - 98\% \\ \hline

Nataraj et al.~\cite{nataraj2011comparative} &	k-NN with Euclidean Distance, System Call-level Monitoring, Forensic Screenshots  & 320 GIST descriptors	& 393/~ \newline 3,131/24 \newline 63,000/531 & Converts binary file to a greyscale image and extracts 320 GIST features to classify malware using k-NN.  &  Accuracy (First Dataset) - 98\% \newline Accuracy (Second Dataset) - 97\% \newline Accuracy (Third Dataset) - 72\%  \\ \hline

Kancherla et al.~\cite{kancherla2013image} & SVM, Gabor Filter, Wavelet Transformation & 2-dimensional vector of file size & 25,000/6 & Converts a binary file to a 2D vector, extracts Gabor, Intensity and Wavelet features, and then uses SVM for malware detection and comparison against each feature set. & Accuracy (Gabor) - 93.23\% \newline Accuracy (Lightweight) - 94.32\%\\  
\hline

Ahmadi et al.~\cite{ahmadi2016novel} & Forward Stepwise Selection Algorithm, XGBoost Classifier & Hex byte as a greyscale pixel & 21,741/9 & Calculates set of features from different categories of hex and assembly view of PE and applies feature fusion algorithm to generate the most effective concatenation of features categories such as Haralick features and Local Binary Patterns features to classify a malware. & Accuracy (with Img. Feat.) - 95.5\% \newline Accuracy (comb. of other feat.)- 99.40\% \\ 
\hline

    
Zhao et al.~\cite{xiaolin2018research} & Homology Analysis, B2M, Imporved SIFT, Guassian Pyramid, DBSCAN & 66 byte feature descriptor & 1,789/11 & Extracts features from binary sample, creates texture fingerprint from the features, and then applies clustering to group the texture fingerprint of malware into families. Afterwards, the homology analysis was performed to classify the malicious unknown samples. & Accuracy - 77.6\% \\ 
\hline

Masouleh et al.~\cite{silva2018improving} & HOG, AE, Summary Statistics Features from icon pixels, HDBSCAN, K-NN & Icon Pixels of varied length & 1,138/~ & Detects PE malware by extracting HOG, AE and summary statistics features form icon pixels, clusters them, and then trains machine learning classifiers. & Accuracy: 84.4\% \newline TPR: 80.2\% \\ \hline

Su et al.~\cite{MD:Kywe2015} & Colour Correlogram Algorithm, Cosine Similarity  & App Icons and Screenshots & 30,625 crawled mobile apps & Detects camouflaged apps on mobile market places using indexed based text and image retrieval systems. & FPR: 9.22\% \\ \hline

\hline 
\multicolumn{5}{c}{\textbf{Image Hashing-based Methods}} \\ \hline
\hline

Xiaofong et al.~\cite{xiaofang2014malware} & LSH SURF, Euclidean Distance & 64*64 image vector & 8,410/25 & Computes a feature fingerprint and applies LSH to return the top most visually (structurally) similar variants for malware classification.  & Precsion/Recall/F-Score - 80\% to 90\% \\
\hline

Malisa et al.~\cite{malisa2016mobile} & LSH & Mobile app screenshots & 150,000 mobile apps & Detects impersonation attacks by capturing run-time app activities screenshots and measures similarity through LSH.  & FPR - 15\% \newline Impersonated Apps- 43,904 \\ \hline

Andow et al.~\cite{andow2016study} & Fuzzy Hashing & App icons & 2,500 mobile apps & Detects imposter apps through text analytics and static program analysis, where later measures the similarity of app icons through fuzzy hashing. & Manual App Review Reduction: 977 apps out of which 22 app icons are same \\ \hline

Jiao et al.~\cite{MD:Jiao2015} & pHash & App images & Not Given & Detects repackaged apps by comparing hash codes and if the the similarity exceeds a threshold, then flag the app as repackaged app. & Not Given \\ \hline

Long et al.~\cite{MD:Long2014} & wrestool, FLANN index, L2 distance & 1024 dimensional vector & Not Given & Measures similarity between apps by extracting images, converts images to features, and then uses L2 distance for similarity checking. & Recall - 83\% \newline Precision - 85\% \\ \hline

\hline 
\multicolumn{5}{c}{\textbf{Neural Network-based Methods}} \\ \hline
\hline

Zhang et al.~\cite{zhang2016irmd} & Opcode Sequence, Histogram, Dilution, Erosion, CNN & 2-Dimensional opcode images & 17,808/10 & Disassembles binary executables into opcodes sequences, converts the opcodes into images and then compares with the known malware sample using CNN. & Accuracy (Max.) - 96.7\% \newline TPR (Max.) - 94.3\% \newline TNR (Max.) - 99.1\%  \\ 
\hline

Kim et al.~\cite{kim2017malware} & GANS, t-SNE Algorithm & Average of all the rows and columns of the data &  9,970/9 & Utilises transfer learning method to pre-train the generator for generating malware images close to real malware for the possible detection of zero-day attacks. & Accuracy (Avg.) - 96.39\% \\
\hline

Kim et al.~\cite{kim2018zero} & DAE, GANS & 63*135 pixel image & 10,800/9 & Devises an architecture by combining several deep learning methods of DAE, GAN, and transfer learning to construct a malware detector that is robust to noise and zero-day attacks. & Accuracy (Avg.) - 96.74\%  \\ 
\hline

Kim et al.~\cite{kim2018detecting} & GANS, VAE & Image size 256*128 \newline Latent Space Size - 270 dimension & 10,867/9 & Generates malware from latent space based features and helps the detector to learn features with unseen observations. & Accuracy (avg.) - 96.97\% \\
\hline

Ni et al.~\cite{ni2018malware} & SimHash, CNN, Bilinear Interpolation & 8*8, 8*16, 16*16, 16*32, 24*32, 28*32 &  10,805/9 & Extracts opcode sequence based features, uses SimHash method to calculate similarity, converts SimHash values to greyscale images and then uses CNN to identify malware families. & Accuracy (Avg.) - 98.86\% \\
\hline

Yan et al.~\cite{yan2018detecting} & CNN, LSTM, Truccated back-propagation Algorithm & 64*64 greyscale image & 21,736/9 & Uses CNN for extracting image based features and LSTM to learn opcode sequences features for malware detection.  & Accracy - 99.88\% \\
\hline

\hline

Cui et al.~\cite{cui2018detection} & CNN, Bats & Multiple representations depending on original file size & 9,342/25 & Malicious code is converted into greyscale image and then images are classified using CNN having convolution and subsampling layers. & Accuracy (Avg.) - 94.5\% \\
\hline

Yue et al.~\cite{MD:Yue2017} & CNN and weighted softmax loss & VGG-verydeep-19 CNN model with 60 layers & 9,342/25 & Proposes a solution to overcome class imbalance problem in malware classification through CNN of 43 and two dropout layers. & Accuracy - 98.63\% \\
\hline

Le at al.~\cite{le2018deep} & CNN, CNN-UniLSTM, CNN-BiLSTM & 10000 byte file size & 10,868/9 & Utilises three different CNNs to improve the performance of malware classification. & Accuracy(CNN) - 95.8\% \newline Accuracy (CNN-UniLSTM) -  98.12\% \newline Accuracy(CNN-BiLSTM) - 98.20\% \\
 \hline
 
Kalash et al.~\cite{kalash2018malware} & CNN & 320 GIST descriptors & Malimg - 9,339/25 \newline MS - 1,868/9 & Converts malware binaries to images and trains CNN for classification. & Accuracy (Malimg) - 98.52\% \newline Accuracy (Microsoft) - 99.97\% \\
 \hline

Jiawei et al.~\cite{su2018lightweight} & CNN & 64*64 greyscale image & 500/4 & Utilises CNN to detect and classify IoT DDOS malware residing in IoT devices. & Accuracy (2-class) - 94.0\% \newline Accuracy (3-class) - 81.8\% \\
 \hline
 
Cao et al.~\cite{cao2018efficient} & CNN & 2-D array with specified widths 28*28 / 56*56 & Vision Lab and Microsoft Datasets/20 & Works for online environment where a browser plug-in sends suspected malicious file to detection server running CNN for further analysis. & Accuracy - 95\% \\
 \hline 
 
Akrash et al.~\cite{akarsh2019detailed} & CNN-LSTM & Greyscale images of varying size & 9,389 malware samples & Detects malware families using CNN-LSTM model and tests against various parameters. & Accuracy: 95\% to 97\% \\ \hline
 
Venkatraman et al.~\cite{venkatraman2019hybrid} & CNN, CNN BiLSTM, CNN BiGRU & Greyscale images of varied length & 52,000 malware samples collected from private and public sources & Detects and classifies known and unknown malware samples through a hybrid deep learning and image analysis techniques. & Accuracy: 94.8\% to 96.3\% \newline F1 Score: 90.3\% to 91.6\% \\ \hline

Darshan et al.~\cite{sl2019windows} & CNN, n-gram features & Greyscale images of varying size selected by feature selector & Self-Generated Dataset: 400 samples \newline Malheur Dataset: 7,433 samples & Detects and classifies known and unknown malware PE files by executing them at run-time, extracting n-gram features, converting features into images and feeding into CNN. & Accuracy: 97.96\% \newline F-Score, Precision, Recall: 0.97 \\ \hline

Xue at al.~\cite{xue2019malware} & CNN, n-grams & Greyscale images & 174607 malware samples & Classifies malware using probability scoring and machine learning models with the purpose to overcome the drawbacks of static and dynamic malware analysis methods i.e. code obfuscation and computation cost. & Accuracy: 98.82\% \\ \hline

Rajasegaran et al.~\cite{rajasegaran2019multi} & VGGNet & App icons & 1.2 million apps from google Playstore & Detects counterfeit apps by using a pre-trained VGGNET that combines content and style embeddings. & Precsion: 8\% to 12\% \newline Recall: 14\% to 18\% \newline Counetrfeit Apps: 7,246 from the top-10,000 apps \\ \hline
\end{tabular}
\end{table*}


\subsection{Summary of computer vision based malware detection:}

In this section, we surveyed research work that used computer vision methods to detect and classify malware more accurately and efficiently. The main motivation of using computer vision methods is to overcome the limitations of static and dynamic code analysis methods, where the former suffers from the problem of code obfuscation and detecting new malware types, and the latter is computationally expensive. The discussed computer vision-based research work has shown promising results in terms of detection and classification accuracy, and in some cases, offers high accuracy against the detection of obfuscated or encrypted malware. Moreover, some of the research work in this area is independent of the malware execution environment and computationally inexpensive.  

In computer vision-based malware detection, the main idea is usually to convert malware binaries or behaviour of malware to an image and then perform analysis on the image using methods such as feature extraction, image comparison, and image hashing. Similar to phishing detection, we categorise the malware detection approaches to \emph{\bfseries{i) image feature-based  methods}}, \emph{\bfseries{ii) image hashing-based methods}}, and \emph{\bfseries{iii) neural networks-based  methods}}. In addition, we also found some work under the \emph{\bfseries{image representation-based methods}} that visualise malware activities such as API calls, resources consumption, file structure decomposition, and frequency of API calls, as images and then compares the unknown sample with abnormal actions using image processing techniques such as bitmap converter, thread graph, treemaps, and colour mapping. We summarise all the work we surveyed in Table~\ref{tab2:Mal_summary}.





Next, we discuss observed challenges and limitations over all the work, with the purpose to make the contribution of computer vision more effective in the area of malware detection, and thereby deploying solutions commercially. Our study on existing computer-vision malware detection methods reveal various findings:





\begin{itemize}
    \item \textbf{Malware Dataset and Families:} We found that most of the work used Microsoft Malware Classification Challenge (BIG 2015) as their dataset. This dataset is 0.5 terabytes in size, consisting of disassembly and bytecode of more than 20K malware samples. Another malware dataset which is most often used is the Malimg dataset. This dataset contains 9,339 malware samples from 25 different malware families. We also found a work proposed by~\cite{trinius2009visual}  that has set-up honeypots in a controlled network environment for real-time malware detection. 
    
    We also observe that most of the work did not validate their work against multiple datasets. For instance, Silva et al.~\cite{silva2018improving} evaluated their method only on the BIG dataset, whereas, sl et al.~\cite{sl2019windows} used Malheur dataset for evaluation. Nevertheless, there are methods such as Le et al.~\cite{le2018deep} and Kalash et al.~\cite{kalash2018malware}, that performed evaluation on more than one datasets. We also notice that most of the work was evaluated on the specific set of malware families provided by the datasets and discussed the extension of their work to new malware variants. There are very few works that perform regressive testing on a broad set of malware families. For example, we found work by Han et al.~\cite{han2015malware} that has performed their experiments on 50 different malware families. To build a robust malware detection system, it is necessary to collect latest and large datasets, containing several malware families and their corresponding variants, such as EMBER~\cite{ember}.  Moreover, malware datasets should contain samples from a range of platform to check the diversity of malware detection method.  An example here is Winnti Windows malware family that was first explored in 2013, however, recently (in 2019) its Linux variant was discovered which dated back to 2015~\cite{winnti}. 




   \item \textbf{CNN and GANS:} We noticed that understandably CNNs are an emerging approach in detecting and classifying malware efficiently. We found that several recent works have used different versions of CNNs, either to detect malware presence or to classify malware types. CNN-based methods are fed either with greyscale or coloured malware images and then tested with a number of different input and output layers and various parameters. As a matter of fact, each of the research work in neural network category used a different number of inputs, middle, and output layers and also tuned with different hyper-parameters. There is a need to investigate for a generic neural network model that has a capacity to tune layer and parameters for high malware detection accuracy automatically. 
Recently, researchers also used GANS to generate new malware training data to detect zero-day attacks~\cite{kim2018detecting, kim2017malware, kim2018zero}. However, the research is still underway as it involves many challenges. For instance, it is still not clear that malware generated through GANS resembles the real malware and how much helpful they are in the detection of zero-day attacks.

  \item \textbf{Real world adoptions:} The computer vision-based malware detection has not fully adopted by the industry, despite a significant amount of work happening. We did not find any company that explicitly mentions the usage of such work in their commercial products. One possible reason for this gap is that computer vision-based malware detection methods are not mature enough to be deployed as commercial products. The proposed methods neither have convincing results nor cover all aspects of malware detection. For instance, most of the methods were tested on a specific dataset and a particular network environment. These methods have not been deployed in open network to verify their robustness, reliability, and performance at large scale. Moreover, there is no standard benchmark to compare these work against each other. 
  
  


    \end{itemize}

%% file: Section_05.tex
\section{Visual Similarity based Traffic Anomaly Detection}\label{sec:ADA}

Computer vision methods also alleviate the limitations of traditional traffic anomaly detection approaches that are usually based on statistical features of traffic flows. For instance, a conventional approach that detects short and long-lasting anomalies through variances in the number of bytes (traffic volume) cannot identify low-rate attacks~\cite{kuzmanovic2003low}. An example of a low-rate attack is port scanning that does not consume much bandwidth when it tries to access an abnormally large number of ports on a single host~\cite{AD:Fontugne2008}. Also, such approaches cannot identify attacks that are statistically close to global traffic behaviour. Finally, these traditional approaches are unable to detect zero-day attacks. As a result, much effort has been put into detecting anomalies through computer vision methods.


In this section, we survey computer vision-based traffic anomaly detection approaches and categorises them based on their method of detection. We found work fitting into the categories of \textit{\textbf{i) image representation-based methods, ii) image feature-based methods,}} and \textit{\textbf{iii) neural network-based methods.}} However, we highlight the volume of work we found for computer vision-based traffic anomaly detection is much less compared to \textit{phishing detection} and \textit{malware detection}. A possible reason for this gap is that traffic anomaly detection is a highly non-trivial application for image processing techniques, as compared to phished web pages or malware binaries that can be directly represented as images
We also did not find any work that utilised \textit{image hashing-based methods} to detect traffic anomalies. We further discuss the possible reasons behind this at the end of the Section~\ref{sec5:discussion}.

\subsection{Image representation-based methods}
\label{sec5:img}
\begin{figure*}[!h]
\centering
\includegraphics[width=0.9\textwidth, keepaspectratio]{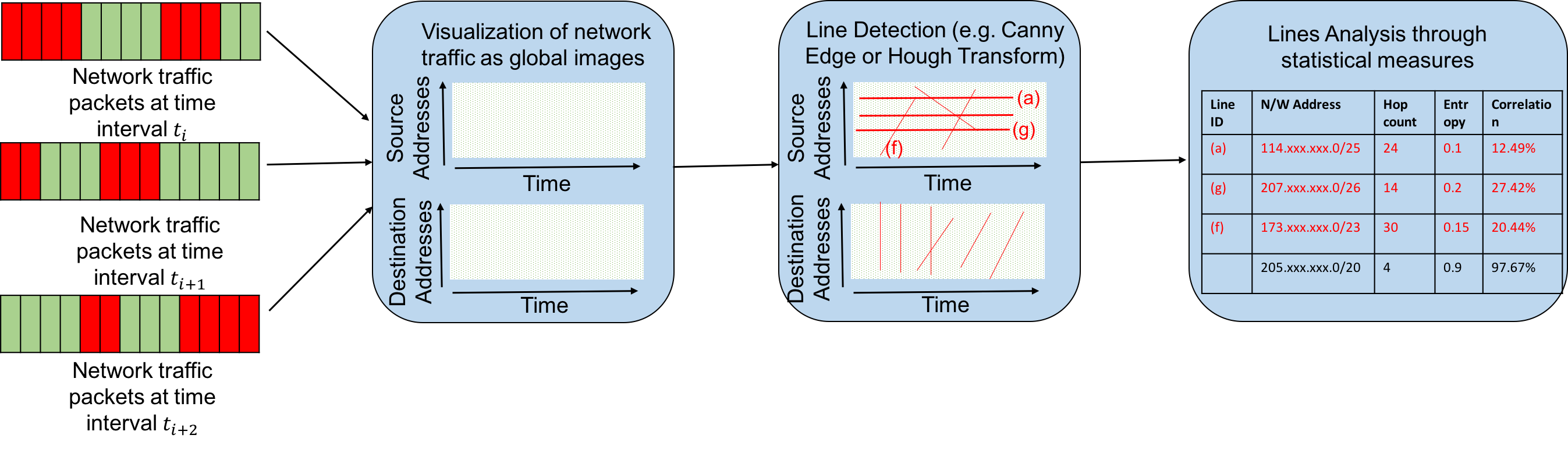}
\caption{Generic Process of Anomalous Traffic Detection using image-based representation methods}
\label{fig:traffic_example}
\end{figure*}

Similar to Section~\ref{sec4a}, the image representation-based methods detect traffic anomalies by visualising traffic data or logs as images and then analysing them using computer vision algorithms. The traffic data usually consists of network packets with header and payload sections, where header contains information such as source IP address, destination IP address, source port, destination port, and protocol names. Studies have shown that network traffic can have strong patterns of behaviour over several timescales~\cite{barford2002signal}. Therefore, by passively monitoring packet headers of network traffic at regular intervals and then generating images of traffic usage (which can be correlated with previous images), can help to detect anomalous behaviours. For example, in case of anomalous traffic such as DoS attack, the usage pattern of network at time $t_i$ is different from previous normal states $t_1, t_2, ....t_{i-1}$.  These usage patterns at different time intervals can be represented as visual images; for example, images of traffic volume originating from source or destination IP addresses, traffic flow between IP addresses, or intensity of traffic flows. The images are then analysed using computer vision algorithms (e.g. Hough Transformation~\cite{duda1971use}, Canny Edge Detection Algorithm~\cite{canny1986computational}, and Video Compression Algorithms~\cite{wiegand2003overview}), to find peculiarities or further characterise the anomalies into several categories (e.g. random attacks, targeted attacks, multi-source attack, and port scans). Figure~\ref{fig:traffic_example} shows a generic process of detecting anomalous traffic using image-based representation methods.

Kim et al.~\cite{AD:Kim2005_netviewer1, AD:Kim2005} is one of the early work in detecting anomalous traffic using image representation methods. Authors proposed an approach, \textit{Netviewer}, that used image processing and compression techniques for detecting and identifying network attacks in real-time.  First, samples of network traffic are converted into images. The traffic samples are packet header traces that consist of data such as source/destination addresses, port numbers, and traffic volume in bytes. For instance, an image may be represented as traffic volume in bytes originating from a source IP address, or the traffic between a source and destination IP addresses/port numbers.  Second, the network traffic images, from regular time intervals, are then considered as video frames such that a sudden change in a scene (traffic pattern) indicates a possible traffic anomaly. To analyse the data in video frames, various techniques from video compression and image processing such as Discrete Cosine Transform (DCT) coefficients and absolute difference of image pixels are applied. For instance, a mean square error (MSE) is calculated between video frames (which are collected at regular interval) to detect anomalies. Finally, to identify attacks and victims, line or edge detection algorithms are used. For example, consider a representation where the x-axis of an image represents destination IP addresses, and the y-axis represents the source IP addresses. A horizontal line in this image, is an indication that a source is accessing multiple destinations, for example, during worm propagation. A horizontal line may also indicate a host scan of destination machines by a single source with a high probability. Similarly, a vertical line indicates that several sources are accessing a single destination (e.g. a DDoS/Botnet attack emanating from multiple machines towards a single destination). Moreover, Netviewer also estimated the movement of attack patterns using motion prediction algorithms.


Authors used real traffic traces from three major networks to evaluate their approach. The networks were University of Southern California (USC), Korea Research Environment Open NETwork (KREONet2), and Texas A\&M University campus~\cite{hussain2003framework}. Authors claimed that their method has an average accuracy of 92.3\% and a false positive rate of 0.3\% when the source IP address was considered in the packet header. There was a 5\% reduction in accuracy and 0.45\% increment in false positive rate when only the destination address was considered in the analysis. Authors also compared their approach with popular Intrusion Detection System (IDS) Snort and showed that the results of both systems agree with each other. However, both systems differ from their methodology perspective. By investigating payload and packet header, Snort is capable of identifying the source of malicious activities as well as functions performed by those malicious activities. On the other hand, Netviewer investigates packet headers only and reports suspicious IP addresses and the pattern of abnormality in aggregation manner (e.g. by observing the number of packets arriving at a specific destination IP addresses at time $t$). This makes Snort superior in presenting granularity details of attacks. Similarly, Snort employs qualitative analysis (rule-based), whereas Netviewer is quantitative (aggregation-based). As a result, Snort may miss the identification of some unusual traffic if the rules are not adequately defined. For example, heavy traffic flows from a single host in a network may not be detected by Snort without an operational rule. In contrast, Netviewer can identify such flows as anomalous because of its aggregation-based method.





In~\cite{AD:Kim2006}, authors used a similar approach to Netviewer~\cite{AD:Kim2005_netviewer1}, with the main difference being using image pixel intensities and discrete cosine transform (DCT) coefficients to detect scene changes in video frames. In addition to other packet header data, an image also represents the intensity of traffic of corresponding IP address through colour and darkness of each pixel. For example, a packet count from a source or destination IP address can be represented as the intensity of the pixel in the image representation. The anomalies are then detected using real-time and postmortem analysis, where the former exploits the variance of pixels intensities of a traffic image and the latter employs the variance of 8-by-8 DCT of the traffic image. Similar to~\cite{AD:Kim2005_netviewer1}, the approach was tested on the traces from three major networks and showed that the real-time analysis has a 0.22\% and 2\% higher false positive and accuracy rate respectively compared to postmortem analysis based on DCT coefficients.


Besides, the authors also compared their the real-time analysis method with classical detection theory-based \textit{Neyman-Pearson test} and showed the false alarm rate of 0.37\% and the detection rate of 74.6\%. As compared to the previous approach by Kim et al.~\cite{AD:Kim2005_netviewer1}, the results of this approach are slightly better, which indicates that the real-time analysis (image pixels intensities) on the network traffic is a better performer for the anomaly detection than the postmortem analysis. Moreover, the advantage of using real-time analysis is that it is effective against flood types of attacks. When images are defined as the intensity of traffic flow in the source address, destination address, or destination port, and analyzed as video frames, then the distribution of abnormal traffic flows in some frames are expected to be much different from normal and historical distribution. 




Above two approaches~\cite{AD:Kim2005_netviewer1, AD:Kim2006} detect anomalies by representing traffic flow as images and then using a threshold to determine significant changes from the images of regular network traffic. However, these methods may not be able to detect several other anomalies that do not rely on traffic flow intensities. To this end, Fontugne et al.~\cite{AD:Fontugne2008} proposed an approach that is capable of detecting unusual traffic behaviours by pointing out the abnormal distribution of traffic features. In this approach, network traffic is mapped to snapshots, and then pattern recognition technique such as Hough Transform is used to identify anomalies. Pattern recognition not only allows for unsupervised learning, but also enable the detection of ambiguous short-lived traffic. To detect anomalies, first, time-based sliding windows are adjusted, and then network traces in each sliding window are converted into images using four traffic features (i.e. source address, destination address and ports). Once converted, the Hough transformation is applied to detect lines in snapshots representing unusual and excessive use of traffic features.  Once a line is found, network traffic data involved in those lines is retrieved and summarised as an event. An event provides detailed information such as the percentage of use of all protocols, the entropy of each traffic feature, timestamps, and IP addresses, about packets extracted from the whole traffic flow. Moreover, these events are clustered together based on destination or source address to detect the anomaly.



The approach was tested on the traffic traces from the Measurement and Analysis on the Wide Internet (MAWI) archive~\cite{sony2000traffic}, which were collected in 2004. Overall, it detected 73.8\% of anomalies from a dataset containing 2000 anomaly traffic and 630 normal traffic. Moreover, the proposed approach was compared with a statistical approach non-Gaussian multi-timescale models~\cite{dewaele2007extracting} and showed efficient detection of short and long-term anomalous traffic. The proposed method~\cite{AD:Fontugne2008} detected 297 more anomalies which were not identified by a statistical method. Also, the method claims to have shorter detection delay as compared to the statistical method (i.e. this approach detects traffic anomalies within three seconds, whereas the statistical method needs at least one minute to detect anomalies). Though the proposed approach is capable of detecting more anomalies than the compared statistical approach, the authors did not discuss false positive anomalies, thus requiring further analysis. In addition, authors applied their method to an old dataset which may not reflect the current characteristics of network traffic. Hence, there is a need to validate this work with the latest network traffic datasets such as CAIDA IPv4, IPv6, or KeroNet. All in all, the approach is capable of detecting volume-based attacks through traffic features. 





Jeong et al.~\cite{AD:Jeong2010} proposed a hierarchical approach that performed image analysis in two-tiers to detect network attacks. The first-tier detects random attacks (e.g. DDoS attacks) by analysing global images whereas the second tier detects semi-random attacks (e.g. post scanning attacks) by examining the local traffic images. Global images are created based on the network traffic such that X-axis represents the destination IP addresses and Y-axis represents the source IP addresses. The intensity of the image indicates the traffic flow between the source and destination pairs. Similar to~\cite{AD:Fontugne2008}, lines in the images are identified using the Hough transform. These lines are an indication of high usage of a port number, IP address or a high traffic flow between specific source and destination addresses. Based on the information provided by the lines in the global image, local images are created that contain more details about the attack. Local images show traffic volume for each destination host and port number in a network domain. Once created, Connected Component Labelling (CCL)~\cite{he2017connected} is applied to local images to detect traces of mostly used destination ports (a.k.a post scanning). Authors evaluated their approach on Netflow data of KREONet, which collected network traces from July 20 to July 27, 2003. The method analysed 60,000 to 70,000 events in 3 seconds, thus showing efficiency. However, the authors did not discuss the accuracy of detecting network attacks using the proposed method. 


Similar to~\cite{AD:Jeong2010}, Kim et al.~\cite{kim2016defending} also represent unusual patterns of traffic as lines on a 2D image space. However, their approach detects traffic anomalies that generate only from a DDoS attack with IP spoofing. The unusual patterns which represent DDoS attack sources are recognised using Canny Edge Detection Algorithm. Afterwards, signature-based pattern extraction algorithm, \textit{pivoted movement}, is applied to prevent DDoS attacks. The pivoted movement algorithm correlates information such as IP address and media access control (MAC) pairings.  



The overall proposed scheme~\cite{kim2016defending} works as follow: First, the incoming packets to a victim’s machine are converted into a 2D image using the network tool, NetSCENE, introduced by the authors. NetSCENE monitors the network packets at regular intervals and analyzes the aggregated data. The Canny Edge Detection algorithm is then applied to an image to identify DDoS attack packets with subnet spoofing. The patterns of DDoS attack appear as edges on NetSCENE where each edge is linked to the local network addresses at which targets (i.e. effected machines under the control of attacker) are located. Once lines are identified, the victims’ machine alerts the local source router about those network addresses which then applies the pivoted movement algorithm by paring the source IP address with MAC address. The reason to pair the IP address with the MAC is that legitimate packets pair one IP address with one MAC address whereas, the attack packets use different IP addresses with one MAC address. This is often called pivoting the IP addresses.  Moreover, the two-step detection process also reduces the risk of false positives and false negatives. Finally, the illegitimate IP addresses are marked with signatures and are filtered out. To test the approach, the authors used two different datasets; IPv4 Routed/24 and IPv6 topology datasets from CAIDA~\cite{caida1, caida2}. Results showed an average false positive rate of 1\% and an average false negative rate of 0.5\% for canny edge detection and pivoted movement algorithm, respectively. 



While image representation-based methods offer promising results, there is also a limited amount of work that extracted image-related features from traffic data and then analysed those features for possible traffic anomaly detection. Image-feature based methods help in the discovery of abnormalities in the network that does not depend on traffic volume and would otherwise go undetected by image representation-based methods. For instance, web attacks such as SQL injection can be detected through statistical features (e.g. length of URL, string input types, string matching), which are difficult to detect through image representation-based methods. Moreover, by exploiting the correlation among image related features of network traffic, it is possible to identify the type and characteristics of anomalies precisely and at the same time, reduce false positive and increase true positive rate. We next discuss image feature-based methods that enable the detection of traffic anomalies in the network.





\subsection{Image feature-based methods}

In comparison to image representation-based methods, we did not find a significant amount of work in image feature-based methods for traffic anomaly detection. The most plausible reason being, usually different image features capture different elements of the image such as key-points, colour distribution, and intensity levels and it is not straight forward to decide which feature might be important to identify traffic anomalies with high accuracy. Therefore, such methods may not provide additional benefits over statistical feature-based conventional methods. The two work we found belonging to this category, extracted features from a traffic sample/pattern and determined the presence of an anomaly in network traffic. The extracted features mainly include entropy, contrast, pixel randomness, and local homogeneity of the image.






Early work in this area is by Tan et al.~\cite{AD:Tan2015} that detected DoS attacks using features of network traffic. The overall approach consists of three steps. In the first step, basic features such as Timestamp, Source IP, Destination IP, Source port, source bytes, average duration of host, average duration of all services, number of ICMP packets, number of packets to all services, and number of SYN flags, are generated from network traffic packets captured at the destination network. In the second step, dimensionality reduction using Principle Component Analysis (PCA)  is performed on the generated features. Finally, the selected lower-dimensional features are used in training and testing phase for possible detection of DoS attacks. In a training phase, standard profiles are generated for various types of legitimate/normal traffic records (i.e., TCP, UDP and ICMP traffic). These profiles are built based on the subspace feature set provided in step two. The normal profiles are then converted into two-dimensional feature matrix (image), named Triangle Area Maps (TAM),  that reveals the hidden correlations between the features of network traffic. To do so, Multivariate Correlation Analysis (MCA)~\cite{tan2013system} is applied that selects most discriminative features. The TAM is then stored in a database along with mean and standard deviation of the earth mover’s distances between individual training samples and the mean of the given training samples. In the testing phase, images (TAM) of individually tested records are generated and compared against the normal images from the Training Phase using the EMD algorithm.


The method was evaluated using two datasets; KDD Cup 99 and ISCX 2012~\cite{kdd, shiravi2012toward}. Authors also compared their method with then state-of-the-art detection systems; network intrusion detection system based on covariance feature space~\cite{jin2007network}, triangle- area-based nearest neighbours approach~\cite{tsai2010triangle}, DoS attack detection system using TAM-based MCA~\cite{tan2013system}, and Naïve Bayes based detection approaches~\cite{kumar2013design}. The overall evaluation showed 99.95\% accuracy on KDD Cup datasets, outperforming the techniques in~\cite{tsai2010triangle} and~\cite{jin2007network} by 2.06\% and 7.8\% respectively, and was comparable to the accuracy of~\cite{tan2013system}. On the ISCX 2012 IDS dataset,  the proposed method achieved 90.12\% accuracy outperforming NB-based detection approaches. It also showed promising results in terms of computational complexity by analyzing 59,738 traffic records per second. Another advantage of this method is that it processes each traffic sample individually, which helps in prompt detection of an attack and also increases the probability of correctly classifying a sample than group-based detection mechanisms. Group-based detection mechanisms take a certain number of samples together and then classify them as benign or illegitimate~\cite{jin2007network}. Overall, this work demonstrated to have less detection delay as compared to group-based approaches, yet has equal or better detection precision.



Zou et al.~\cite{AD:Zou2018} proposed a network anomaly detection method that used image processing to analyse network traffic patterns in Cyber-Physical Systems (CPS). A CPS is an integration of physical and software components that interact with each other in different contexts to perform desired functionalities at different spatial and temporal scales~\cite{cps}. Examples of CPS include smart grids, medical monitoring systems, and robotics systems. The proposed method relies on the network throughput of all the devices in a CPS network such that a communication image is created by visualising throughput as pixels. The size of the communication image is equal to the number of CPS devices in the network where each image represents the network resource usage of the whole CPS network at a time sample. The method then applies texture feature analysis~\cite{materka1998texture} on images to extract the spatial information about the network resource usage of the CPS devices. More particularly, Grey Level Co-occurrence Matrix (GLCM) was applied to extract five GLCM texture features which were energy, entropy, contrast, IDM (Inverse Difference Moment), and DM (Directional Moment)~\cite{baraldi1995investigation}. These features measure the intensity of pixel pair repetitions, the randomness of pixels, the extent of a pixel and its neighbours, local homogeneity of the communication image, and alignment of it in terms of the angle, respectively. In order to extract more communication features, a temporal analysis of GLCM features was also carried out. Thus, the spatial and temporal features together determine the communication patterns of the CPS network. Afterwards, a k-NN algorithm is used to detect the anomalies. 

 

The proposed method was experimented on four real-world decentralised applications using Common Open Research Emulator (CORE) simulator~\cite{ahrenholz2010comparison}. CORE is a live emulator that uses existing operating system virtualization techniques to build wired and wireless virtual networks. These networks can be connected to real networks and systems to run various protocols and applications to extend lab test-beds. Authors set up 100 devices on CORE Simulator and run every application for 200 times and capture the network throughput. Authors then simulate the anomaly network traffic by adding different intensity (10\%, 20\%, 30\%) of random network throughput to every device. The overall classification accuracy was 99\%; however, the method was not tested on a real dataset or a real environment. Moreover, the time performance of extracting GLCM features has not been discussed in the paper. 


To summarise,  we only found a limited amount of work under image feature-based methods for traffic anomaly detection. One possible reason may be the difficulty in identifying what features make more sense in traffic anomaly detection, which is potentially solved in the neural-network approaches we next discuss.

\subsection{Neural Network-based methods} 

Similar to the phishing and malware detection, neural-network based approaches have also been explored for traffic anomaly detection. In this category, images are generated from network traffic and then directly fed into neural networks. The network automatically learns latent representations that can be leveraged to detect traffic anomalies.

Wang~\cite{wang2015applications} demonstrated the idea of representing network flow data as images and utilising a deep neural network to detect anomalies. First, the payload bytes from each TCP session is converted to an image, where each byte represents a pixel after normalising to a scale from 0 to 1. Once an image is formed, the images are fed into Multi-Layer Perceptron (MLP) and Stacked Auto-Encoder (SAE) deep learning models for the identification of anomalous protocols in network traffic. The MLP model represents a supervised learning setting (requires labelled data) while the SAE represents an unsupervised setting (can work without labelled data). In the MLP and SAE models, the nodes that have higher significance (low error cost) in the deep layers are selected as features and trained on the data. Classification results show that SAE performs better on traffic classification than the MLP model.



Both MLP and SAE models were evaluated on 0.3 million TCP flow data records collected from an internal network. These records have 58 different protocol types in total while having 17\% of the overall data flows labelled as ‘unknown’. Both methods were able to distinguish 54.9\% of the unidentified protocols by giving a probability of 0.9  to classify a traffic protocol of each network flow correctly. However, the authors provided limited information on the experimental settings (e.g. dataset used) and results of the methods understandably due to the fact that they represent enterprise data.

In an improvement, Li et al.~\cite{AD:Li2017} proposed a CNN based intrusion detection approach while using a graphic conversion approach. The proposed method converts raw packets into an image, and then a CNN is used to learn the features from that graphics. Authors took NSL-KDD dataset~\cite{aggarwal2015analysis} as a case study for the evaluation of their approach.  In the first step, 41 features, that contain integer or float features, symbolic features and binary features, were extracted from each sample of NS-KDD dataset. The extracted features are then converted into binary vectors with 64 dimensions. Authors used one-hot encoder mapping to convert symbolic features into binary vectors and standard min-max scaling to convert other continuous features into binary. The binary vectors are then converted into $8\times 8$ grey-scale pixel images and input to a CNN model. 

Authors used ResNet50 and GoogLeNet as the CNN architectures, where former was trained with 100 epochs, and 256 batch size and latter was trained with 100 epochs and 64 batch size. Authors reported that both the models showed accuracies ranging from 79\% to 82\% in detecting anomalies in NSL-KDD dataset. Moreover, the authors compared their approach with traditional classifiers such as Random Forest and SVM and showed that CNN performed better. However, the authors did not report the detailed analysis of anomalies (e.g. type and pattern related to each anomaly). Also, the paper did not mention about the treatment of unknown data types/packets by CNN and the effect of such unknown data packets on the performance of the approach. 


Akin to above, Wang et al.~\cite{wang2018using} also proposed an approach based on representation learning where network flow is converted into images and fed into a CNN classifier. Authors used the dataset released by the University of New South Wales (UNSW), referred to as NB-15~\cite{unswdata}. This dataset contains 25,400,44 network communication data records, and the total size of the dataset is 100GB. Compared to the KDD dataset, NB-15 has a large volume of data and more accurate ground truth labelling for anomalous traffic (i.e. attack types). Authors, pre-processed the dataset by performing traffic split, traffic mark, fixed-length interception, and format conversion task. In traffic split, the original traffic file is divided into small traffic files according to the session flow where a session is a communication record of the same IP 5-tuple. The 5-tuple includes the source IP address, source port, protocol type, destination IP address and destination port. Afterwards, the session units are marked (traffic mark) by finding the corresponding entry in a CSV file using IP 5 tuple and the traffic generation time.  Once labelled, each byte of session unit data converts into an integer ranging from 0-256 to create a fixed-length interception. The integers then transform into a greyscale image of 784 (28 $\times$ 28) bytes and feed into the CNN that consists of two convolution layers and two pooling layers. The CNN was trained using cross-entropy loss with a learning rate of 0.001 and a batch size of 50 over 1,000 epochs. The authors reported accuracy of 97.3\% and an F1 measure of 0.985. However, there was no comparison made with traditional classifiers or with the KDD-dataset. As such, it is difficult to establish where this method stands compared to both classical anomaly detection methods as well as computer vision-based methods.

Though different from traffic anomaly detection methods, Internet traffic classification is another related area where computer vision methods have been applied. Zhou et al.~\cite{zhou2018classification} classified botnet encrypted and unencrypted traffic by using CNN-based features self-learning.  First, the data is pre-processed by dividing the network traffic into sessions such that the size of a session meets the input requirement of the CNN. The sessions are determined through bidirectional flow standard~\cite{anderson2016identifying} where not only incoming but also outgoing packets also participate in malicious activity. To form a session, the authors used five tuples; source IP, source port, destination IP, destination port, and the protocol. After pre-processing, the session data was converted into a greyscale image by normalizing each byte of a session to [0, 1] where one byte corresponds to one pixel. The CNN architecture authors used is a standard set up with two blocks of convolution and pooling layers, followed by a fully connected layer of size 1,024 and a final softmax layer of 12 units.



Authors evaluated their solution on the ISCX-Bot-2014 released by The Canadian Institute for Cybersecurity~\cite{unbdata}. The dataset includes both botnet and normal traffic; however, the authors used only botnet traffic by filtering out botnet communication IP addresses provided by the dataset owner. Results indicated that increasing the training iterations, improves the performance of a method, where precision and recall of Blackhole and SmokeBot botnets achieve more than 85\%. Overall, the technique reached an average accuracy of 99.18\%, which is the highest accuracy achieved among all neural network-based anomaly detection methods. However, there is no baseline comparison of this method with others, as different methods were evaluated on different datasets. 



Similar to~\cite{AD:Zou2018} in image feature-based methods, Moore et al.~\cite{moore2019anomaly} also proposed an intrusion detection system for cyber-physical systems, that maps the controller area network (CAN) data to 2D images. The method extracts network features (e.g. speed and acceleration of a vehicle) of a CAN which are then converted into 2D images. These images are input into CNN for the classification of abnormal traffic. Before image generation, frequency Fourier Transformation is used to perform dimensionality reduction by representing the physical relationships between features. Authors used data from one vehicle network with six different traces. The first five traces were used as labelled data to train a CNN, while the sixth trace was used to test the classification accuracy. Authors reported the accuracy of 93\% after ten epochs.

  
\begin{table*}[!h]
\tiny

\caption{Traffic Anomaly Detection Approaches}
\label{sec5:tab3}
\vspace{-3mm}
\begin{tabular}{p{1.8cm}|p{2.0cm}| p{2.2cm}|p{1.5cm}|p{5.3cm}|p{2.9cm}}

 \hline
\textbf{Ref}	& \textbf{Techniques Applied}	& \textbf{Image Structure} & \textbf{Dataset (Normal/ Anomaly)} &\textbf{Description}& \textbf{Performance Metrics}\\ \hline
    
\hline
\multicolumn{5}{c}{\textbf{Image Representation-based Methods}} \\ \hline
\hline

Kim et al.~\cite{AD:Kim2005_netviewer1, AD:Kim2005} & DCT, Mean Square Error   & $4 \times 4$ DCT coefficients  & 3616/729 & Visualises and detects anomalies in real-time by passively monitoring packet headers and then converts them into video frames using DCT and absolute differences of pixels. & \textbf{Source Address}-\newline Accuracy- 92.3\%, FP- 0.3\%  \newline \textbf{Destination Address}-\newline Accuracy - 87.2\%, FP- 0.75\% \\ \hline
 
Kim et al.~\cite{AD:Kim2006} &	DCT, pixels intensities & $8 \times 8$ DCT coefficients & 3563/ 782 & Visualises and detects anomalies using DCT coefficients and pixel intensities in real-time. & \textbf{Real-time analysis} \newline TP-92.8\%, FP-0.42\% \newline \textbf{Postmortem analysis} \newline TP-94.8\%, FP- 0.20\% \\ \hline

Fontugne et al.~\cite{AD:Fontugne2008} & Hough Transform & Image of network packets with time sliding window & 630/ 2000 & Detects anomalies by converting traffic into images and then applies Hough transform to identify lines representing DoS or port scanning attacks. & Accuracy - 73.8\% \\ \hline

Jeong et al.~\cite{AD:Jeong2010} &	Hough Transform, CCL & Image of network packets at fixed sampling rate & 70,000 events & Detects anomalies using two tier architecture, where first tier detects random attacks such as DoS and second tier detects semi-random attacks such as port scanning. &  Not Given \\ \hline

Kim et al.~\cite{kim2016defending} & Canny Edge Detection, Pivoted Movement Algorithm & 2-Dimensional Image with upper 24 bit and lower 8 bit IPv4/v6 addresses & CAIDA Ipv4/Routed and IPv6 dataset & Detects DDoS attack with IP subnet spoofing using canny edge detection and pivoted algorithm in a self-developed NetSCENE tool. &  FPR: 1\% \newline FNR: 0.5\%\\ \hline

\hline
\multicolumn{5}{c}{\textbf{Image Feature-based Methods}} \\ \hline
\hline

Tan et al.~\cite{AD:Tan2015} &	EMD, PCA, MCA & 2D feature matrix (TAM) & \textbf{KDD cup 99:} \newline 97,260/ 3,892,550 \newline \textbf{ISCX 2012 IDS:} \newline 2,450,329/ 8720	& Detects DoS attacks by extracting image-related network features and then applies EMD algorithm on features.  & \textbf{KDD cup 99:} \newline Accuracy - 99.95\%, FP-1.93\% \newline \textbf{ISCX 2012 IDS:} \newline Accuracy - 90.12\%, FP-7.92\%  \\ \hline

Zou et al.\cite{AD:Zou2018} & GLCM texture features, k-NN, Euclidean Distance & Size of the image is equal to number of CPS devices in the system & 100 CPS device traffic	& Detects anomalies in a CPS by conducting a spatial and temporal analysis of network resources and using textural features. & Accuracy - 99\% \\ \hline

\hline 
\multicolumn{5}{c}{\textbf{Neural Networks-based Methods}} \\ \hline
\hline

Wang et al.~\cite{wang2015applications} & ANN, SAE & 1 byte = 1 pixel & 0.3 million data flow records & Utilises a deep neural network technique to extract and select features significant for detecting unknown and anomalous network protocols in a network traffic. & Unknown Protocol Identification rate - 54.9\% \\
\hline
  
Li et al.~\cite{AD:Li2017} & ResNet50, GoogLeNet as CNN	& $8\times 8$ input image & \textbf{NSL-KDD: } \newline \textbf{Test$^{+}$: }\newline 9711/12833 \newline  \textbf{Test$^{-21}$:} \newline 2152/9698  & Detects intrusions by converting raw packets to images and then input into CNN for automatic feature learning. &  \textbf{ResNet 50:} \newline \textbf{Test$^+$:} Accuracy - 79.14\% \newline \textbf{Test$^{-21}$:} Accuracy - 81.57\% \newline \textbf{GoogleNet:}\newline \textbf{Test$^+$:} Accuracy - 77.04\% \newline \textbf{Test$^{-21}$:} Accuracy - 81.84\% \\ \hline
  

Wang et al.~\cite{wang2018using} & CNN & 78*78 image (784 bytes image) & 100GB (25,400,44 network communication data records) & Converts network flow to images and then feeds into CNN for malicious network identification using representation learning. & Accuracy - 97.3\%, \newline F1 - 98.5\%, \newline Precision - 98.65\%, \newline Recall - 98.4\% \\
\hline

Zhou et al.~\cite{zhou2018classification} & CNN & 1 byte = 1 pixel & 7.93GB network flow data & Classifies botnet encrypted and unencrypted network traffic by using CNN-based features self-learning and images as input. & Accuracy - 99.18\% \\
\hline

Moore et al.~\cite{moore2019anomaly} & CNN & 18x25 pixel image & 6 network traces from one vehicle network & Detects cyber attacks on a CAN bus network by exploiting the features related to physics such as frequency domain. & Accuracy - 93\% \\\hline

\end{tabular}
\end{table*}

\subsection{Summary of computer vision-based traffic anomaly detection:}



\label{sec5:discussion}
In this section, we surveyed existing computer vision-based approaches for detecting network traffic anomalies. The motivation to use computer vision methods, as reported by several papers~\cite{AD:Fontugne2008, AD:Jeong2010, AD:Tan2015}, is to overcome the problem of traditional detection approaches. Most of the conventional approaches are unable to characterise and detect specific types of attacks such low rate attacks, replacement attacks or zero-day attacks. Computer-vision methods help in alleviating such problems by utilising pattern recognition or image processing algorithms such as Hough transformation and Canny Edge Detection, where detection of a single line, edge, or a high-intensity pixel may represent abnormalities. 


We categorised the work on computer vision-based methods under three topics; \textbf{\textit{i) image representation-based methods, ii) image feature-based  methods,}} and \textit{\textbf{iii) neural network based-methods.}} Image representation-based methods monitor network packet headers at regular intervals, generate images of traffic usage, and then correlate/match current network traffic image with previously generated images. The deviated images can be further analyzed to detect attack type by using pattern recognition algorithms mentioned above. Image feature-based methods extract features such as entropy,  contrast,  pixel randomness, local homogeneity, and Triangle Area Maps (TAM), from the network traffic packets and then input into a distance metric or a classifier for anomaly detection. Similar to phishing and malware detection, neural network-based methods in anomaly detection also generates images from network traffic and then fed into a deep neural network. Akin to the above two categories, neural-network is also a recent trend in anomaly detection. We summarise all the work we surveyed in Table~\ref{sec5:tab3}. 




We next discuss some challenges and limitations of existing work in computer vision-based traffic anomaly detection that need to be addressed to make the research more effective and deployable in the network industry.



\begin{itemize}
    \item \textbf{Challenges in Computer Vision-based Anomaly Detection Methods:}  While computer vision methods have extensively been investigated to detect phishing and malware attacks, we find limited work on the detection of network traffic anomalies. As mentioned earlier, one possible reason could be that network traffic is a non-trivial application for computer vision methods. It is not straightforward to convert network traffic flows into images in a meaningful way, and in some cases, may not provide additional benefits compared to statistical anomaly detection methods. Moreover, network traffic uses several protocols such as TCP, UDP, HTTP, and HTTPS, that have different requirement and configurations. Therefore, a generic computer vision solution is harder to implement. Consider an email service provider that wants to detects anomalies such as self-propagating emails which spread through email virus or worms. The anomaly detection approach, in this case, requires images based on features mainly including several connections from a single host, and email header bytes. On the other hand, a video streaming service provider (e.g. Youtube) is more interested in deploying anomaly detection approaches based on traffic volume/flow to its servers. Thus, a generic computer vision-based solution requires further extensive research and testing. This area may further grow with the advances in unsupervised deep learning methods that is currently an active area of study.

    \item \textbf{Detailed Investigation of Anomalous Traffic:} Though existing computer-vision based anomaly detection methods can detect abnormal behaviour in network traffic with high accuracy, we find that existing work did not provide further insights on different types of anomalies such as botnets, DoS, backdoor, worms, and scanning. The methods generalise all these attack types as anomalous behaviours. However, in reality, each of these attack types has different characteristics which can have different impacts. Thus, there is a need to perform a detailed analysis of the detected anomalies that may highlight specific trends and patterns of varying anomaly types.
    
    
    \item \textbf{Traffic Anomaly Datasets:} We found a number of datasets that have been used by existing methods for anomaly detection such as NSL-KDD~\cite{kdd}, KERONet~\cite{hussain2003framework}, CAIDA~\cite{caida1, caida2}, and ISCX~\cite{unbdata}. The majority of work used the NSL-KDD dataset~\cite{kdd}; however, its ability to reflect real-world conditions has been often criticised~\cite{mchugh20001998,brown2009analysis}. The primary concern was the inability to evaluate anomaly detection systems against current and evolving anomalies and network traffic patterns. In this regard, CAIDA~\cite{caida1, caida2} and WIDE~\cite{sony2000traffic} datasets are being periodically updated as compared to KDD dataset. Though, it is always good to have a range of datasets for testing and validation purposes; a benchmark is helpful in setting as a standard. In addition to the standard dataset, continuously evolving datasets with properties such as modifiable, extensible, reproducible, are also required to include new network trends. Nevertheless, we do not find any dataset that serves the purposes stated above. 
    
    


    
    
    \item \textbf{Convolutional Neural Networks (CNN):} We found that the use of CNNs is an emerging trend for detecting anomalous traffic. Several recent work utilised CNNs for network anomaly detection where the image is taken as an input. While CNN is more common neural network model among several works, we also find SAE being used by one method~\cite{wang2015applications}. This direction is expected to grow with further advancements in CNN-based methods.
    
\end{itemize}

%% file: Section_06.tex
 \section{Prototype and Commercial Solutions}
\label{industry}

As we mentioned in the introduction, the use of computer vision methods in network security is not purely academic interest. There is an increasing trend from the industry to use such methods to overcome the limitations of the exiting solutions and obtain a competitive advantage. In this section, we survey computer vision-based tools/software that are offering network security-based services. Tools we study are either developed by individuals working under the capacity of developers or researchers (prototype solutions), or by companies who are trying to improve their existing network security products (commercial solutions). We highlight that this product survey is carried out based on publicly available information. Thus, indeed this list is not comprehensive as many companies might not release the inner workings of their products publicly due to intellectual property reasons. Also, in many cases, the available public information is limited to provide an in-depth analysis of the tools and products.  Our search resulted in 12 such tools/products as enumerated below.  \\ \vspace{-2mm}

\noindent{\textbf{i) Blazar~\cite{blazar}} is a tool offered by Endgame Research~\cite{Endgame} to detect Homoglyph attacks using a Siamese Neural Network. Homoglyph attacks are a type of phishing attacks where the users are prompted to with a ULR that appear visually similar to a legitimate website (e.g. \texttt{www.google.com} vs. \texttt{www.g0ogle.com}). The key idea of Blazar is to convert text URLs to images and feed them to a Siamese Neural Network that can capture visual similarities between characters. This approach works better compared to edit distance based text similarity methods, mainly because of extended character sets. The model was trained using a synthesised dataset that contained pairs of images of popular domain names and modified versions of them that appear visually similar using various extended characters. Blazar also implements a KD tree-based indexing and lookup system to improve performance~\cite{woodbridge2018detecting}. Blazar is released in GitHub as an open-source tool.} \\ \vspace{-2mm}
    
\noindent{\textbf{ii) SpeedGrapher~\cite{Speedgrapher}} is another tool by Endgame Research that detects macro-enabled document-based phishing attacks. In such attacks, the attackers trick victims by requesting them to open malware-embedded documents. The company already had a labelled dataset of macro-based documents that are known to be malware and to develop SpeedGrapher they generated screenshots of such documents using the preview from the Word Interop class~\cite{microsoft}. From the screenshots, a range of features such as prominent colours, blur/blank areas, embedded characters, and icons (using YOLO CNN). Finally, a classification algorithm, Random Forest, is used to detect malicious macros in a document. According to Endgame Research, this is a product that is still under active development and as such no code is released to date.} \\ \vspace{-2mm}
    
\noindent{\textbf{iii) Lookout Phishing AI~\cite{PhishingAI}} is an internet monitoring tool that identifies early signals of a phishing attack and alerts its users of the potential harm.  This tool incorporates a machine learning engine that continuously scans the URLs in a browser and watches the behaviour of the site. Phishing AI claims to use computer vision methods in their website behaviour analysis~\cite{PhishingAI}. However, the exact details are not provided on the company website. Recently Phishing AI was able to detect a large scale phishing campaign targeting non-governmental organizations such as UNICEF~\cite{unicef}.} \\ \vspace{-2mm}
    
\noindent{\textbf{iv) Phish.ai~\cite{phishai}} is a solution that can detect zero-day phishing attacks. Similar to many solutions discussed in Section~\ref{sec:PWA}}, Phish.ai keep a visual signature database extracted from the screenshots of top websites and brands. The API accepts the links as the input and obtain a screenshot of the website and compare its visual signature against ones stored in the database to check whether the new site is trying to impersonate an existing, legitimate site. While it might be interesting to know what types of visual signatures are used in this product, such information is not disclosed by the company. \\ \vspace{-2mm}
    
\noindent{\textbf{v) MalSee~\cite{MalSee}} is a malware detection and classification system that is developed by researchers at Mayachitra Inc. It is similar to the ideas presented in the early work by Nataraj et al.~\cite{MD:Nataraj2011}}, where the malware binaries are represented as images. The tool is offered as a web-based API and claims to provide $\sim$1,000x speedup detection of harmful computer viruses compared to existing methods. \\ \vspace{-2mm}

    
\noindent{\textbf{vi) Anti-Pixm~\cite{Pixm}} by Pixm Inc. is another phishing detection tool that uses deep learning and computer vision to analyze the web pages visually. Compared to other similar products, Anti-Pixm claims to provide real-time, end-device based protection against stealth mode phishing attacks (i.e. the phishing attacks against cloud-based email security tools). When a user clicks a link received in an email, Anti-Pixm browser extension quickly grabs a screenshot in
the background, compares its visual signature with common targets, and blocks the URL within less than a second. The tool claims to find phishing attacks before they are being identified by major cloud-based phishing solutions that maintain blacklists. The exact vision-based deep learning methods used by Pixm is not publicly available.} \\ \vspace{-2mm}
    
\noindent{\textbf{vii) Ironscales's IronShield~\cite{ironscale}} is a cloud-based solution to defend against real-time malware, credential threats, account takeovers, and phishing websites. One specific aspect of the solution is to validate the legitimacy of login blocks using computer vision methods.} \\ \vspace{-2mm}
    
\noindent{\textbf{viii) INKY~\cite{INKY}} is a cloud-based email security solution that blocks spam, malware, and phishing attacks. One of the main features the solution provides is detecting peculiarities and brand forgeries in email graphics and icons. INKY offers seamless integration with Office 365, Microsoft Exchange Solutions, Microsoft Security, and Google G Suit so that phishing emails can be analysed and contained in realtime.} \\ \vspace{-2mm}

\noindent{\textbf{Cyberfish~\cite{Cyberfish}}, \textbf{GreatHorn~\cite{406venture}}, \textbf{Area1~\cite{area1}}, and \textbf{ZeroFox \cite{zerofox}} are similar phishing email detection solutions that use visual signatures. No significant information is available about these solutions apart from the fact that they are using computer vision methods to match visually similar web pages.}

\begin{table*}
\scriptsize
\caption{Computer Vision Based Network Security Products and Prototypes } \vspace{-3mm}
\begin{tabular}{p{2cm}|p{3.7cm}| p{2.3cm}|p{1.5cm}|p{6cm}}

 \hline
\textbf{Product Name}	& \textbf{Security Service} 	& \textbf{Deployment Level} & \textbf{Open Source?} & \textbf{Description} \\ \hline

\hline
   
   Blazar \cite{blazar} & Homoglyph Attacks (Domain/URLs and File Names) Detection &  Browser and Network Infrastructure & Yes	& Utilises Siamese Convolutional Neural Network and Euclidean Distance to measure similarity between original and malicious URLs/file names.\\\hline 
   
   SpeedGrapher \cite{Speedgrapher} & Detects macro-enabled document based phishing & Desktop for Documents e.g. Microsoft Word & No & Utilises various computer vision techniques such as Blur Detection, Blank Detection, Optical Character Recognition (OCR), and Icon Detection to extract features from a document.   \\ \hline
   
   Lookout Phishing AI \cite{PhishingAI} & Detect Phishing Websites & Browser & No & Utilises computer vision techniques to identify the phishing sites by analyzing the use of logos and graphics. \\
   \hline
   
   Phish.ai \cite{phishai} & Detects zero-day phishing attacks & Customized API and web Browser & Yes & Creates a unique and up-to-date computer-vision based database of legitimate websites and afterwards compares a screenshot of suspicious website with the database.\\
   \hline
   
   MalSee \cite{MalSee} & Malware Detection & Web-accessible service & No & Utilises pattern recognition from image processing methods to detect and classify a malware. \\
   \hline
  
  Pixm \cite{Pixm} & Detect Phishing Websites  & Browser & No & Uses computer vision techniques to compare screenshots of websites for possible phishing detection. \\
  \hline
  
  Ironscales' IronShield \cite{ironscale} & Provide multiple security services such as phishing detection, zero-day malware detection, Malicious Email Detection and credential thefts & Cloud level Network Infrastructure, Browser, and Email & No & Utilises computer vision to detect in real-time visual deviations and determines whether or not a login page is legitimate.\\
  \hline
  
  INKY \cite{INKY} & Block spam, malware and phishing attacks & Cloud based network infrastructure, and Email, Browser & No & Uses computer vision to spot differences in email graphics and iconography to detect forgeries. \\
  \hline
  
  Cyberfish \cite{Cyberfish} & Detect phishing attacks in emails and web pages & Cloud and On-Premise Email and Browser & No & Analyzes visual representation of emails and web pages to detect attacks in real-time.  \\
  \hline
  
  GreatHorn \cite{406venture} & Credential Theft Detection & Email & No & Utilises CNN and image representation-based computer vision methods to detect credential thefts via email.\\
  \hline
  
  Area1 \cite{area1} & Phishing detection on multiple platforms & Email, Web and Network & No & Utilises AI and computer vision methods to provide effective phishing protection.
  \\
  \hline
  
  
  ZeroFox \cite{zerofox} & Risk identification & Web/Domain, Social Media Accounts & No & Utilises computer vision techniques like OCR, logo and face detection for risk identification in multiple platforms such as social media, web and domain. \\
  \hline

\end{tabular}
\end{table*}

One key observation regarding commercial tools is that most of the solutions are developed in recent years despite there are published academic work since the mid-2000s as our survey found. One possible reason for this can be the exponential growth in deep learning-driven computer vision methods happened in the last few years making computer vision accessible for other domain experts. Also, such attempts are further exacerbated by the ubiquitous availability of deep learning and computer vision libraries such as Tensorflow, PyTorch, OpenCV, and SciKit Learn that enables rapid prototyping and testing without any specialised knowledge. Thus, we believe that in future, there will be more of these types of security solutions as well as research work. 


Another interesting observation is that apart from Malsee~\cite{MalSee} all the rest of the products we found are built to detect phishing attacks. However, when it comes to academic work, both phishing detection and malware detection had a significant body of work that used computer vision methods. Furthermore, we did not find any commercial or prototype solution that used computer vision in anomaly detection. While there is no conclusive evidence on why this is the case, we believe that from an applied research point of view, using computer vision techniques for phishing detection is conceptually straightforward compared to malware detection or traffic anomaly detection. As computer vision methods further advances, we believe the information security community will start adapting those ideas in malware and traffic anomaly detection as well. Finally, only very few tools were available as open-source tools. One thing that can be done to further enhance the research in this domain is to release the code and the data together with the academic publication.


%% file: Section_07.tex
\section{Open Issues and Research Challenges}
\label{sec:discussion}


Computer vision methods for network security is an evolving research area, and it is likely to grow further with the exponential advances in deep learning-driven computer vision methods. We also provided the evidence that the information security industry has also started to adapt of build security products or features adapting the ideas from computer vision, mainly to address the limitations of existing solutions and to obtain a competitive advantage. As a growing field, there are several exciting research directions and challenges to solve, to build more robust security solutions using computer vision methods. We next discuss such opportunities and challenges. 

\subsection{Hybrid Solutions}
Almost all the work we discussed, used computer vision methods and compared them against traditional methods. We find only a very limited literature~\cite{MD:Han2014, ahmadi2016novel, andow2016study} that propose a hybrid approach of combining conventional and computer vision methods. Nonetheless, it is in favour of security defenders to deploy both types of methods to overcome the limitations of each other.  For example, Fatt et al.~\cite{PW:Fatt2014} used \emph{Favicons} for phishing detection. However, this work does not apply to web pages that do not have \emph{Favicons}. In such a case, traditional text or network-based methods can be merged into this approach. PhishZoo~\cite{PW:Afroz2011} combined image and text-based methods and showed an overall accuracy of 90.2\% in phishing detection.



We find similar examples in malware detection where a single type of method does not holistically cover all aspects of detection and classification. For instance, the method proposed by Nataraj et al.~\cite{MD:Nataraj2011} focused only on the visual representations of a resource section in a malware executable file, which allows an adversary to bypass a detection system if malware exists in other parts of a file. In this case, the use of static-analysis methods (e.g. signature-based) may help in detecting malware that is relocated to different sections of a file. Similarly, the resource limitation in running vision-based dynamic analysis on mobile devices can be addressed by first performing static analysis followed by methods such as~\cite{malisa2016mobile}. Thus, we see the need for more hybrid security solutions that combine traditional techniques with computer vision methods for better reliability and accuracy.

\subsection{Adversarial Example Attacks}
\label{adv:sec}
Several works pointed out that computer vision systems, such as object detection and classification solutions, are vulnerable to adversarial examples. That is, an attacker can craft inputs that can trigger misclassifications, \textit{evasion attacks}~\cite{goodfellow2014explaining,moosavi2016deepfool,kurakin2016adversarial} or feed the training process with adversarial data so that the model is biased towards a direction set by the attacker, \textit{poisoning attacks}~\cite{chen2017targeted,munoz2017towards,shafahi2018poison}. Such attacks have been demonstrated extensively for traditional computer vision or machine learning settings. However, recently multiple works showed such vulnerabilities exist in security solutions.

Example work for \textit{evasion attacks} include Al-Dujaili et al.~\cite{al2018adversarial} and Grosse et al.~\cite{grosse2016adversarial} that demonstrated the generation of functional malware that can act as adversarial examples for model-based malware classifiers. Hu et al.~\cite{hu2017generating} proposed MalGAN, a GAN-based generative model capable of generating malware samples that can bypass classifiers such as random forest and logistic regression. Rosenberg et al.~\cite{rosenberg2018query} is another comparable work.

Recent work~\cite{rigaki2018bringing} also showed how malware traffic could be synthesised to bypass intrusion detection systems in practical scenarios. Authors tried to mimic the behaviour of Facebook chat traffic by feeding its network flow parameters to GAN for a predefined number of epochs. The trained GAN generator then communicates its output parameters to the command and control server so that malware can change its traffic accordingly misguiding middleboxes to think malware traffic as Facebook chat traffic.




Biggio et al.~\cite{biggio2011bagging} demonstrated \textit{poisoning attacks} on spam filtering and intrusion detection scenarios where an adversary is capable of controlling a subset of samples that are used to train or update a classifier. The attacker carefully constructs these samples such that the classifier can be misguided. For instance, in spam filtering attack, adversaries can modify spam emails by adding some non-suspicious words which are likely to appear in legitimate emails whereas, in intrusion detection, adversaries may inject poisoning pattern in the network that matches with legitimate activities. 

As vision-based network security solutions become more and more mainstream, attackers will inevitably be exploring adversarial attacks. These attacks are exacerbated by the easy access to software libraries that can generate such attacks (e.g.~\cite{papernot2016cleverhans}). As our survey indicated, at this stage, the research focus is more on getting working systems with the required levels of accuracy. None of the work we presented in Section~\ref{sec:PWA},~\ref{sec:MDA}, and~\ref{sec:ADA} looked into the attacks against their systems. Assessing the adversarial robustness of classifiers and coming up with defence strategies is an active area of research in computer vision and machine learning~\cite{xu2019adversarial, akhtar2018threat}. We believe that in the next phase of research in vision-based network security must look into the robustness of classifiers by exploring the defensive mechanisms proposed in computer visions systems such as adversarial training, model distillation, feature squeezing, gradient hiding, and blocking transferability~\cite{chakraborty2018adversarial} and adapt those to network security solutions.

\subsection{One-shot and Few-shot Learning}



Most of the methods that we discussed in the above sections used deep learning models such as CNN that are well known for their high training data requirements. In object classification or image recognition, it is easy to collect a large volume of data. However, collecting an equivalent amount of data from security events is somewhat tricky. For instance, in the case of malware and traffic anomalies, it is not practical to obtain large volumes of samples as these events are rare and often go unnoticed. Moreover, malware samples and security incident data (e.g. logs and configurations)  are usually not publicly released for safety reasons as well as to protect corporate secrets. Finally, to build labelled datasets for security problems, specialised domain experts are required compared to image labelling or voice transcribing tasks. Although crowdsourcing has been tried (e.g. PhishTank), the ensuring quality remains challenging. Hence it is necessary to explore solutions that require lesser volumes of labelled data.

A possible solution to this problem is the ideas of one/few-shot learning~\cite{vinyals2016matching,snell2017prototypical, qiao2018few, gidaris2018dynamic, woodward2017active, socher2013zero, fei2006one, xian2017zero} and unsupervised feature learning/self-supervised learning~\cite{henaff2019data,he2019momentum} where models learn from minimal or no labelled data. This is an active area of research in computer vision. However, its application in vision-based security solutions remains mostly unexplored. There have been recent attempts in classifying malware through few/one-shot learning models~\cite{hsiao2019malware, tran2019image, atapour2019kings}. Tran et al.~\cite{tran2019image} proposed a  few-shot learning approach to malware classification by first converting malware binaries into greyscale images and then adapting meta-learning models (i.e. matching networks and prototypical networks). The results showed an average accuracy of 92.4\% with a prototypical network for1-shot learning and 95.3\% accuracy with 4-shot learning. Similarly, Hsiao et al.~\cite{hsiao2019malware} used one-shot Siamese neural network to classify unknown malware images. Atapour-Abarghouei et al.~\cite{atapour2019kings} also proposed a ransomware classification approach using one-shot learning through data augmentation. Nevertheless, few-shot learning is still an emerging area in cybersecurity, and extensive research must be conducted to understand how such methods can be effectively utilised.

\subsection{Open Set for Classification}

Majority of the work we discussed operates under a closed set assumption. For example, a malware classifier will classify a given sample as one of the known malware families or as benign. If a sample from a new malware family appears, the classifier will still make a similar decision. Similarly, the same behaviour will happen for a new benign sample as well. In fact, in such settings, there is a possibility that the classifier is making very confident yet highly inaccurate decision~\cite{scheirer2012toward, geng2018recent} as the classifier is not trained to handle \textit{unknown} (i.e. data samples that are out of the distribution of the training set) data samples~\cite{boult2019learning}.


While classifiers that can't handle unknown inputs are sufficient for prototyping, in real-world deployments, often there will be inputs that are entirely outside the distributions of the training data the classifier has seen. This situation may not be critical in some applications like photo tagging. However, when it comes to network security, the ramifications of such solutions can be disastrous. This problem also becomes challenging because of the incomplete knowledge of the world during training (i.e. only the known classes are accessible). Thus, steps must be included to make sure the model knows its limitation on the closed set operation. None of the work we surveyed addressed this challenge, and we believe as the vision-based network security solutions become more mainstream. this is an area that should show growth.

An emerging solution to correctly identify known and unknown samples is through \textit{open-set recognition}. Open set recognition describes a scenario where new classes that were not seen in the training phase appear in testing and requires the classifiers to not only accurately classify known classes but also effectively deal with unknown ones. In the most basic sense, classes utilised in testing are not present in training ~\cite{scheirer2012toward}. This is currently an active area of research in computer vision~\cite{dhamija2018reducing, raj2019improving, oza2019c2ae, liu2019large, yoshihashi2019classification, hein2019relu}. For instance, Dhamija et al.~\cite{dhamija2018reducing, raj2019improving} introduced a novel loss function and regularisation methods to improve the handling of background and unknown inputs. Oza et al.~\cite{oza2019c2ae} proposed an approach based class conditioned auto-encoders and using the reconstruction loss for open set recognition. Similarly, Liu et al.~\cite{liu2019large} introduced Open Long-Tailed Recognition (OLTR) that handles imbalanced classification, few-shot learning, and open-set recognition.


Some recent work in network security has started looking into the open set classification. Cruz et al.~\cite{cruz2017open} detected intrusions in the network by considering both closed set and open set recognition problem. Authors proposed a fine-grained recognition approach that considers all intrusion detection problem as a recognition problem and categorically assigns intrusion detection as a closed set problem and intrusion recognition as an open set problem. Authors used a loosely open set dataset which is labelled according to individual intrusion types (e.g. (sendmail, snmp guess), rather than more general attack categories (e.g., DoS). To validate the approach, two different classifier types are employed - Gaussian RBF kernel SVMs, which are not theoretically guaranteed to bound open space risk, and W-SVMs, which are theoretically guaranteed to bound open space risk. Results showed that W-SVM offers better performance in classifying unknown classes. 

Similarly, Henrydoss et al.~\cite{henrydoss2017incremental} proposed an approach that recognises unknown intrusion classes by understanding the characteristics and also performs incremental learning for updating the classifiers using feedback from the classifier itself after successfully identifying the unknown classes during query time. Authors used a multi-class classifier that is proposed explicitly for open set recognition setting and utilised Extreme Value Machine (EVM) to classify intrusion detection data. The EVM has a capability to perform kernel-free, nonlinear, variable bandwidth outlier detection in combination with incremental learning.  We recommend interested readers to look into a survey paper from Rudd et al.~\cite{rudd2016survey} for open set recognition approaches in intrusion detection. Incorporating open-set recognition techniques to network security solutions is vital, and we are still at the early stages of this research direction. We believe this is as an open yet challenging area that needs a significant amount of further work.

\subsection{Online Learning and Real-time Predictions}

Many of the work that we surveyed were operating on offline mode where a data collection phase followed by a training phase. In the real-world, often network security systems must deal with real-time data, rapid fluctuations, and trends. Also, models need to be re-trained frequently with the availability of new data. As such, when models are deployed, there must be mechanisms to update them at short intervals using methods such as online learning~\cite{bottou2004large, sahoo2017online}. However, there are many challenges associated with such set-ups. Catastrophic forgetting~\cite{mccloskey1989catastrophic, ratcliff1990connectionist} is one of the challenges where the learning model might forget its prior knowledge when fine-tuned with new data.  This is widely believed to be a serious problem for neural networks~\cite{goodfellow2013empirical}. Also, in online learning, designs must consider poisoning attacks as well.

Another aspect of handling real-time data, especially related to traffic anomaly detection is how to do traffic flow sampling accurately. In large networks, it is not practical to analyse every traffic flow. As a result, it might be essential to explore the work proposed in traffic sampling~\cite{tune2011sampling,liu2016one,yang2018elastic,tune2014ofss} and integrate them into the training pipeline. 

%% file: Section_08.tex
\section{Conclusion}
\label{sec:conclusion}

Network security threats are becoming increasingly common as well as sophisticated to a point where not only traditional security solutions but also machine learning-based security solutions are becoming less effective. As an alternative, academic research and commercial solutions providers are exploring the feasibility of using methods from computer vision in combination with machine learning models. We surveyed such work under three broad topics; \textit{phishing detection}, \textit{malware detection}, and \textit{traffic anomaly detection}. Overall our survey showed that there are distinct advantages of using computer vision methods, especially when it comes to detecting zero-day attacks as well as developing more scalable and accurate phishing detection systems. Also, our survey found that in order to stem further research in this area, it is necessary to establish much larger and more recent security datasets that can be conveniently used to establish benchmarks and compare different solutions. Finally, we discussed the potential research directions in this domain such as hybrid solutions combining traditional methods with computer vision methods, building solutions that are resilient of adversarial attacks, few-shot learning to address the challenges in obtaining large samples of data, and open set classification which is required when it comes to real world deployments. 
